\documentclass{article}
\date{}
\usepackage{multirow,color, url, graphics}
\usepackage{subfigure}
\usepackage{float}
\usepackage{epstopdf}
\usepackage{listings}
\usepackage{xcolor}
\usepackage{amsmath}

\usepackage{geometry}
\usepackage{multicol,graphicx,amssymb,mathrsfs,amsmath,pifont,amscd,latexsym,amsthm,color,fancyhdr,CJK,url,hyperref,longtable, pifont, subfigure,epstopdf,setspace,multirow,colortbl,tabularx,threeparttable, booktabs, verbatim, bbm,listings, balance}
\usepackage[linesnumbered,boxed,algosection]{algorithm2e}

\newtheorem{Definition}{Definition}
\newcommand{\tabincell}[2]{\begin{tabular}{@{}#1@{}}#2\end{tabular}}

\begin{document}
\lstset{
basicstyle=\ttfamily,
columns=fullflexible,
showstringspaces=false,
keywordstyle= \color{ blue!70},commentstyle=\color{red!50!green!50!blue!50},
frame=shadowbox,
rulesepcolor= \color{ red!20!green!20!blue!20}
}

\title{Automating Release of Deep Link APIs for Android Applications}
\author{Yun Ma, Ziniu Hu, Dian Yang, and Xuanzhe Liu}

\maketitle
\begin{abstract}
Unlike the Web where each web page has a global URL to reach, a specific ``content page'' inside a mobile app cannot be opened unless the user explores the app with several operations from the landing page. Recently, deep links have been advocated by major companies to enable targeting and opening a specific page of an app externally with an accessible uniform resource identifier (URI). To empirically investigate the state of the practice on adopting deep links, in this article, we present the largest empirical study of deep links over 20,000 Android apps, and find that deep links do not get wide adoption among current Android apps, and non-trivial manual efforts are required for app developers to support deep links. To address such an issue, we propose the Aladdin approach and supporting tool to release deep links to access arbitrary location of existing apps. Aladdin instantiates our novel cooperative framework to synergically combine static analysis and dynamic analysis while minimally engaging developers to provide inputs to the framework for automation, without requiring any coding efforts or additional deployment efforts. We evaluate Aladdin with popular apps and demonstrate its effectiveness and performance.
\end{abstract}

\section{Introduction}
One key factor leading to the great success of the Web is that there are hyperlinks to access web pages and even to specific pieces of ``\textit{deep}'' web contents. For example, the hyperlink \textit{\url{https://en.wikipedia.org/wiki/World_Wide_Web\#History}} points to the ``\textit{history}'' anchor location of the ``\textit{WorldWideWeb}'' wiki page. With hyperlinks, the users can directly navigate from one web page to another and thus reach arbitrary locations on the Web with just click-through behaviors. Indeed, the hyperlinks play a fundamental role on the Web in various aspects, e.g., enabling users to navigate among web pages and add bookmarks to interested contents, and making search engines capable of crawling the web contents~\cite{uLink:MobiSys2016}.

In the current era of mobile computing, it was reported that the Internet usage from mobile devices has already overtaken that from desktops~\cite{mobileInternet}. Accordingly, the apps have been the dominant entrance to access the Internet compared to web pages. However, mobile apps historically lack the considerable support of hyperlinks. At the early age of mobile apps, developers focus on only the features and functionalities, without having strong motivations to provide hyperlink-like support to a specific in-app content. Accessing a specific in-app content requires users to launch this app and land on the ``home'' page, locate the page/location containing the content by a series of actions such as search-and-tap and copy-and-paste, and finally reach the target. Compared to the Web, the support for such ``\textit{hyperlinks}'' is inherently missing in mobile apps so that users have to perform tedious and trivial actions. Other advantages from traditional Web hyperlinks are naturally missing as well.

Realizing such a limitation, the concept of ``\textbf{Deep Link}'' has been proposed to enable directly opening a specific page/location inside an app from the outside of this app by means of a uniform resource identifier (URI)~\cite{deeplink}. Intuitively, deep links are ``hyperlinks'' for mobile apps. For example, with the deep link ``\textit{\url{android-app://org.wikipedia/http/en.m.wikipedia.org/wiki/World_Wide_Web}}'', users can directly open the page of ``\textit{WorldWideWeb}'' in the Wikipedia app. Currently, various major service providers such as Google~\cite{GoogleAppIndexing}, Facebook~\cite{FacebookAppLinks}, Baidu~\cite{BaiduAppLink}, and Microsoft~\cite{BingAppLinking}, have strongly advocated the concept of deep links, and major mobile OS platforms such as iOS~\cite{universallinks} and Android~\cite{applinks} have encouraged their developers to release deep links in their apps. With deep links, mobile users can automatically navigate from one app to a specific page/location of other apps installed on their devices, without manually switching apps.

Indeed, deep links bring various benefits to current stakeholders in the mobile computing ecosystem. Mobile users can have better experiences of consuming in-app content. For instance, they can click through the deep links and thus directly navigate to the desirable target app pages, without having to spend too much time and traverse too many intermediate app pages. App developers can open their deep links to others who are interested in the content, data, or functionality of their apps, so that they can find potential collaborators to realize the ``composition'' of apps. As a result, the ``\textit{page visit}'' and traffic of apps can increase and even explore potential revenues. However, it is unclear how deep links have been supported so far in the state of the broad practice.
%Although it was reported~\cite{uLink:MobiSys2016} that deep links have not been widely adopted by Android apps, there exists no extensive or comprehensive study of deep links .
To address such a question, this paper first conducts an empirical study and uncovers the following findings:

\begin{itemize}
\item \textbf{Increasing amount of deep links with app-version evolution}. Comparing the first version and latest version of the top 200 apps from Wandoujia \footnote{Wandoujia now owns over 3.5 million users and 2 million Android apps. See its website via \url{http://www.wandoujia.com}}, the percentage of apps that support deep links increase from 35\% to 87\%;
\item \textbf{Low-coverage of deep links of current apps}. Although it is observed that the number of deep links keeps increasing, the current coverage of deep links are rather low. More specifically, 73\% of the top 20,000 apps from Wandoujia do not have deep links, and 18\% of the apps have only one deep link;
\item \textbf{Non-trivial developer's efforts.} We make in-depth study of deep-link support at source-code level. By carefully checking open-source Android apps from GitHub that have explicit commit information related to implementing deep links, developers need to manually modify 45--411 lines of code to implement one deep link. Such a finding indicates that supporting deep links requires non-trivial developer efforts.
\end{itemize}

Essentially, releasing deep links for apps is to support programmable app execution to reach specific locations. However, manually implementing such programmable executions is non-trivial for released apps due to the high complexity of current apps. One possible solution is to leverage program analysis of apps to extract execution traces. However, static analysis cannot reach the dynamic contents generated at runtime, and the state-of-the-art dynamic analysis tools for Android apps suffer from achieving low code coverage~\cite{choudhary2015automated}.

To address such a challenge, in this paper, we propose \textbf{\textit{Aladdin}}, a novel approach that can help developers efficiently \textbf{a}utomate the re\textbf{l}ease of \textbf{A}ndroi\textbf{d} app's \textbf{d}eep l\textbf{in}ks  based on a cooperative framework. Our cooperative framework combines static analysis and dynamic analysis as well as engaging minimal human efforts to provide inputs to the framework for automation. In particular, given the source code of an app, Aladdin first uses static analysis to find how to reach activities (each of which is the basic UI component of Android apps) inside an app most efficiently from the entrance of the app. After developers select activities where deep links to dynamic locations are needed, Aladdin then performs dynamic analysis to find how to reach fragments (each of which is a part of a UI component of Android apps) inside each activity. Finally, Aladdin synthesizes the analysis results and generates the templates that record the scripts of how to reach arbitrary locations inside the app. Aladdin provides a deep-link proxy integrated with the source code of the app to take over the deep-link processing, and thus does not instrument the original business logic of the app. Such a proxy can accept the access via released deep links from third-party apps or services. We evaluate Aladdin on some popular Android apps. The evaluation results show that Aladdin can cover a large portion of an app, and the runtime overhead is quite marginal.

To the best of our knowledge, Aladdin is the first work to automate the release of deep links of Android apps without any intrusion of their normal functionality, and thus establishes the foundation of ``\textit{web-like}'' user experiences for mobile apps. More specifically, this paper makes the following major contributions.
\begin{itemize}
    \item{We conduct an extensive empirical study of current deep links based on 20,000 popular Android apps, uncovering the current status of deep links and the developer efforts to implement deep links.}
    \item{We propose an approach to automating the release of deep links based on a cooperative framework, with only very minimal developers' configuration efforts and no interference of the normal functionalities of an app.}
    \item{We evaluate the feasibility and efficiency of our approach on popular Android apps.}
\end{itemize}

The rest of this paper is organized as follows. Section~\ref{sec:background} describes the background of Android apps and deep links. Section~\ref{sec:motivation} illustrates the developer efforts when adding deep links for Android apps. Section~\ref{empirical} presents the empirical study of current deep links. Section~\ref{sec:approach} describes our approach to automating the release of deep links. Section~\ref{sec:implementation} shows the implementation details. Section~\ref{sec:evaluation} evaluates our approach with real-world apps. Section~\ref{sec:discussion} discusses some issues and extensions of our approach. Section~\ref{sec:related} highlights related work and Section~\ref{sec:conclusion} ends the paper with concluding remarks and future work.

\section{Background}\label{sec:background}
Essentially, deep links enable the direct access to arbitrary locations inside mobile apps, providing the experience analogous to hyperlinks that enable the direct access to any location of web pages. In this section, we give some background knowledge of Android apps and deep links.

\subsection{Conceptual Comparison of Android Apps and Web}
\begin{figure}[t]
\centering
  \includegraphics[width=0.6\textwidth]{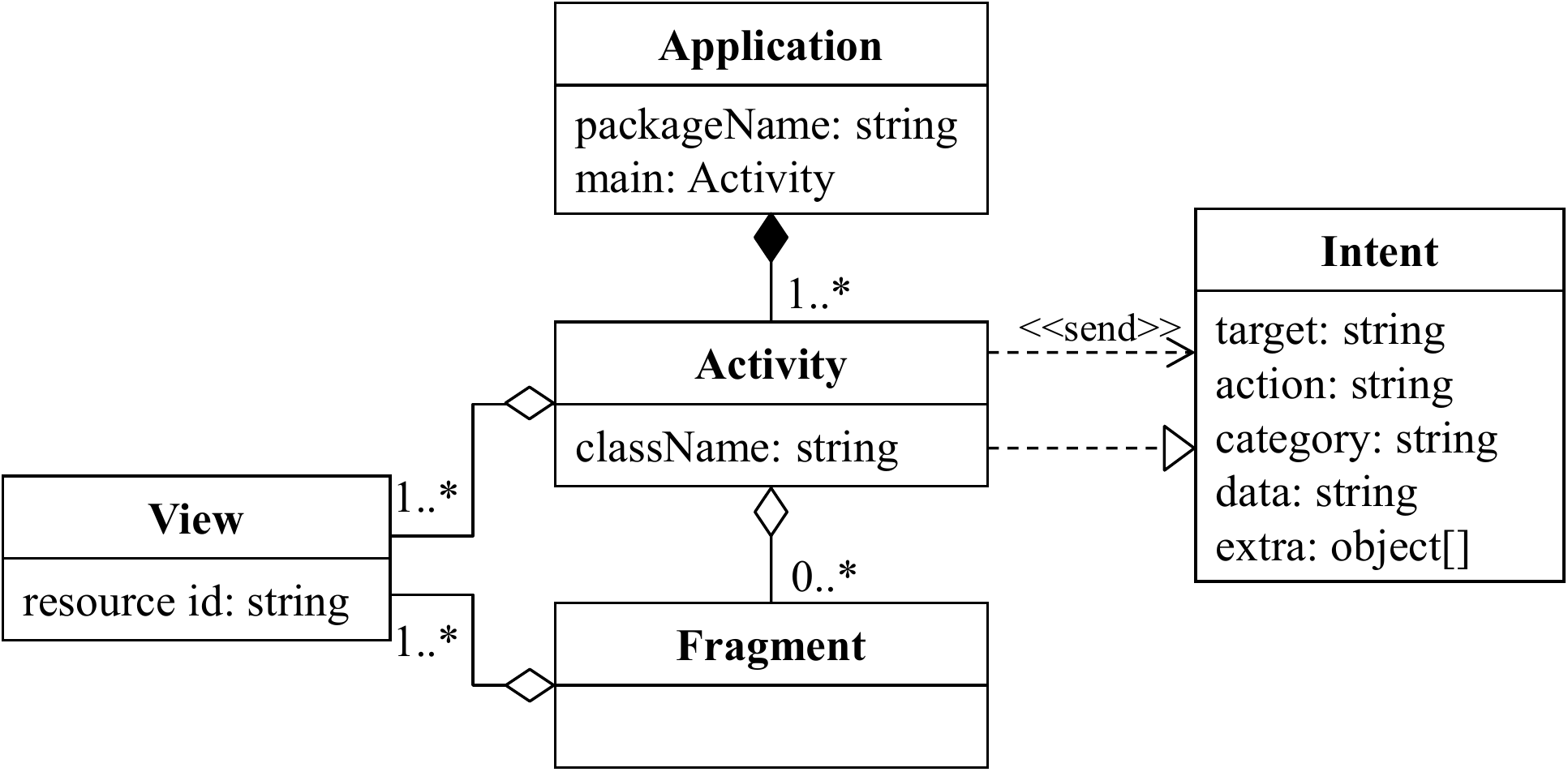}
  \caption{Android application model.}~\label{fig:activity}
\end{figure}
In Figure~\ref{fig:activity}, we illustrate the structure of a typical Android app from Android developer guide~\cite{AndroidGuide}. An app, identified by its package name, usually consists of multiple \texttt{Activities} that are loosely bound to each other. An activity is a component that provides a user interface with which users can interact and perform some tasks, such as dialing phones, watching videos, reading news, or viewing maps. Each activity is assigned a window to draw its graphical user interface. One activity in an app is specified as the ``main'' activity, which is first presented to the user when the app is launched.

For ease of understanding, we can draw an analogy between Android apps and the Web, as compared in Table~\ref{table:concepts}. An Android app can be regarded as a website where the package name of the app is like the domain of the website. Therefore, activities can be regarded as web pages because both of them are basic blocks for apps and websites, respectively, providing user interfaces for users to interact with. The main activity is just like the home page of a website.

An activity has several views to display its user interface, such as \texttt{TextView, ButtonView, ListView}. Views are similar to web elements consisting of a web page, such as \texttt{<p>}, \texttt{<button>}, and \texttt{<ul>}. When a web page is complex, frames are often used to embed some web elements for better organization. Frames are subpages of a web page. Similarly, since the screen size of mobile devices is rather limited, Android provides \texttt{Fragment} as a portion of user interfaces in an activity. Each fragment forms a subpage of an activity.

Activities and fragments hold the contents inside apps, just like web pages and frames encapsulate contents on the Web. In the rest of this paper, we use the term ``page'' and ``activity'' exchangeably, as well as ``subpage'' and ``fragment'' exchangeably, for ease of presentation.

Transitions between activities are executed through \texttt{Intents}. An intent is a messaging object used to request an action from another component, and thus essentially supports Inter-Process Communication (IPC) at the OS level. Note that intents can be used to transit between activities from both the same apps and two different apps. There are two types of intent: (1) \textit{explicit} intents, which specify the target activity by its class name; (2) \textit{implicit} intents, which declare \texttt{action}, \texttt{category}, and/or \texttt{data} that can be handled by another activity. Messages are encapsulated in the \texttt{extra} field of an intent. When an activity $P$ sends out an intent $I$, the Android system finds the target activity $Q$ that can handle $I$, and then loads $Q$, achieving the transition from $P$ to $Q$.

\begin{table}[t]
\centering%\small
\caption{Conceptual comparison between Android apps and the Web.}\label{table:concepts}
\begin{tabular}{l|l}
 \hline
  \textbf{Concepts of Android Apps} & \textbf{Concepts of Web}\\
  \hline
  app & website\\
  package name & domain\\
  activity & web page\\
  main activity & home page\\
  view & web element\\
  fragment & frame\\
  \hline
\end{tabular}
\end{table}

\subsection{Deep Links}\label{sec:deeplink}
The idea of deep links for mobile apps originates from the prevalence of hyperlinks on the Web. Every single web page has a globally unique identifier, namely URL. In this way, web pages are accessible directly from anywhere by means of URLs. Typically, web users can enter the URL of a web page in the address bar of web browsers, and click the ``Go'' button to open the target web page. They can also click through hyperlinks on one web page to navigate directly to other web pages. Indeed, the hyperlinks are instrumental to many fundamental user experiences such as navigating from one web page to another, bookmarking a page, or sharing it with others, e.g., like \texttt{del.icio.us}\footnote{Del.icio.us. \url{https://del.icio.us/}}.

Compared to web pages, the mechanism of hyperlinks is historically missing for mobile apps. A page of an app has only an internal location. Accessing a page in an app has to be started by launching the app, and users may need to walk through various pages to reach the target one. For example, a user can find a favourite restaurant in a food app such as \texttt{Yelp}. Next time when the user wants to check the restaurant information, she has to launch the app, search the restaurant again, and then reach the page of the restaurant's details. In other words, there is no way for the user to directly open the restaurant page even if she has visited before.

To solve the problem, deep links are proposed to enable directly opening a specific page inside an app from the outside of this app with a uniform resource identifier (URI). A deep link can launch an already installed app on a user's mobile device (similar to loading the home page of a website) or it can directly open a specific location within the app (similar to deep linking to an arbitrary web page in a website)~\cite{uLink:MobiSys2016}. The biggest benefit of deep links is not limited in enabling users to directly navigate to a specific location of an app with a dedicated link, but further supports other apps or services (e.g., search engines) to be capable of accessing the internal data of an app and thus enables ``communication" of apps to explore more features, user experiences, and even revenues. Today, all major mobile platforms, including Android, iOS, and Windows, have made various efforts to support mobile deep links.

\begin{figure}[t!]
\centering
    \subfigure[Google App Indexing]{
    \label{fig:appindexing} %% label for first subfigure
    \includegraphics[width=0.4\columnwidth]{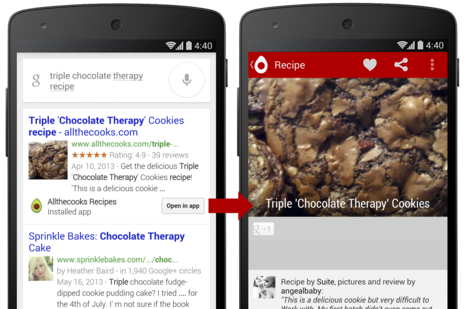}}
    \subfigure[Facebook App Links]{
    \label{fig:applinks} %% label for second subfigure
    \includegraphics[width=0.45\columnwidth]{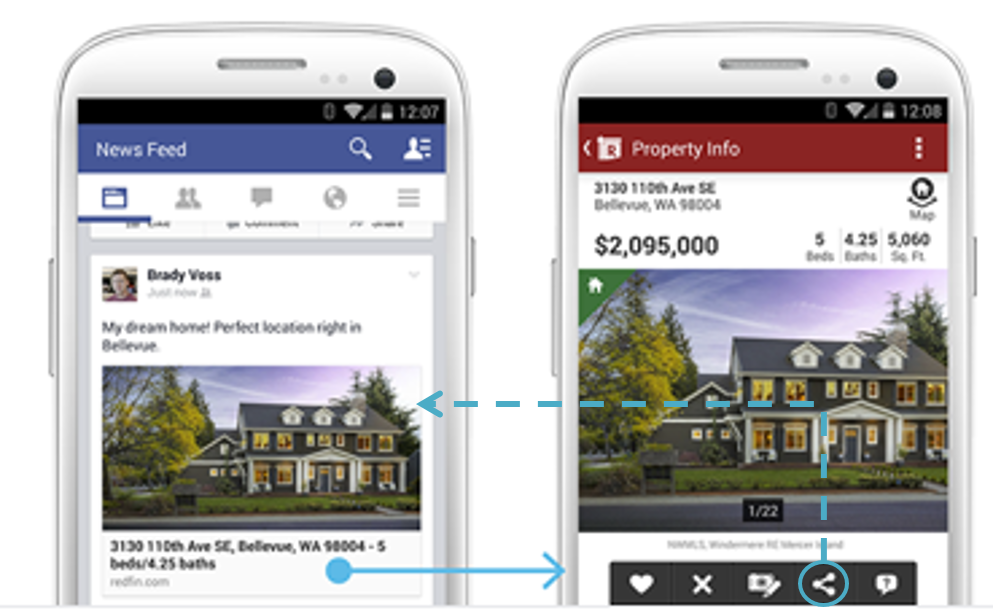}}
  \caption{Usage scenarios of deep links.}~\label{fig:example}
\end{figure}

There are several usage scenarios of deep links on mobile devices. Here we show several industrial practices.
\begin{itemize}
\item {\textbf{In-App Search}. In-app search enables mobile users to search contents inside apps and enter directly into the app page containing the information from search results. For example, Google proposes app indexing~\cite{GoogleAppIndexing} to realize in-app search. When a search result can be served from an app, users who have installed the app can go directly to the page containing the result. Figure~\ref{fig:appindexing} shows an example of how app indexing works for mobile users. When a user searches a term ``\textit{triple chocolate therapy recipe}'' on Google, one result comes from the `\textit{`All the cooks Recipes}'' app. If the user clicks the button ``Open in app'', then the target app page of the search result is directly opened.}
\item {\textbf{Bookmarking}. Mobile users can create bookmarks to the information or functionalities inside apps for later use~\cite{uLink:MobiSys2016}. For example, a flashlight app has a fantastic effect that its users may frequently use. So the app could provide a deep link to the functionality of the fantastic effect and enable users to create a bookmark on their home screen. Afterwards, users could directly activate the fantastic effect by just clicking a bookmark on the home screen.}
\item {\textbf{Content Sharing}. With deep links, mobile users can share pages from one app to friends in other social networking apps. For example, Facebook designs app links~\cite{FacebookAppLinks} as an open standard for deep links to contents in third-party apps. Figure~\ref{fig:applinks} shows an example of app links. When someone uses the \texttt{airbnb}, he/she can share the content to \texttt{Facebook} (the dashed line) with a link that makes it possible to jump back into the \texttt{airbnb} from that piece of content (the solid line).}
\item \textbf{App Mashup}. Other than simple content sharing, deep links can further act as the support for realizing ``\textit{app mashup}''~\cite{Ma:ICWS2015,TSC15Liu,TSC15Huang,TSC16Huang}. For example, a user books a hotel and views restaurant using \texttt{Expedia} and \texttt{Yelp}, respectively. Suppose a deep link can share the location information of hotels from \texttt{Expedia} to map the information of restaurants in \texttt{Yelp}. Then, it is possible to automatically create a trip itinerary by mashing up hotel and restaurant information that the user viewed and interacted with, through clicks on the deep link with system support such as Appstract~\cite{Suman:Mobicom16}. In fact, it is also observed that the ``Now on Tap'' service~\cite{nowontap} from recently released Google Pixel smartphone has leveraged deep links and app indexing to realize such app-mashup user experiences.
\end{itemize}

\section{Motivating Example}\label{sec:motivation}
In practice, deep links are implemented based on the mechanism of Inter-Process Communication provided by mobile systems. For Android, current deep links are located to activities and implemented based on implicit intents. For activities to be deep linked, developers first add intent filters when declaring activities in the \texttt{AndroidManifest.xml} file. Intent filters specify the URI patterns that the app can handle from inbound links. When an implicit intent is sent out, the Android system launches the corresponding activity whose intent filter can match to the intent, thus realizing the execution of a deep link. The code snippet in Listing~\ref{code:intentfilter} illustrates an example of intent filter for links pointing to \url{http://www.examplepetstore.com}. In order to enable the deep links to be opened within Web browsers, the \texttt{action} element should be \texttt{android.intent.action.VIEW} and the \texttt{category} element should include \texttt{android.intent.} \texttt{category.BROWSABLE}.

\begin{scriptsize}
\begin{lstlisting}[caption={Intent Filter Configuration of Deep Links}, label = {code:intentfilter},
	language=XML]
<activity android:name="com.example.android.PetstoreActivity">
 <intent-filter>
  <action android:name="android.intent.action.VIEW"/>
  <category android:name="android.intent.category.DEFAULT"/>
  <category android:name="android.intent.category.BROWSABLE"/>
  <data android:scheme="http"/>
  <data android:host="www.examplepetstore.com"/>
 </intent-filter>
</activity>
\end{lstlisting}
\end{scriptsize}

After declaring the intent filter, the next step is to add the business logic to handle the intent filters. The code snippet in Listing~\ref{code:intentfilterlogic} shows the logic of handling the intent filter defined previously. Once the Android system starts the activity through an intent filter, the business logic is called to use the data provided by the intent to determine the app's view response. The methods \texttt{getDataString}() can be used to retrieve the data associated with the incoming intent during the launching callbacks of the activity such as \texttt{onCreate}() or \texttt{onNewIntent}().

\begin{scriptsize}
\begin{lstlisting}[caption={The Logic of Handling the Intent Filter}, label = {code:intentfilterlogic}, language=Java]
protected void onNewIntent(Intent intent) {
   String action = intent.getAction();
   String data = intent.getDataString();
   if (Intent.ACTION_VIEW.equals(action) && data != null) {
      String productId;
      productId = data.substring(data.lastIndexOf("/") + 1);
      Uri contentUri = PetstoreContentProvider
                .CONTENT_URI.buildUpon()
                .appendPath(productId).build();
      showItem(contentUri);
   }
}
\end{lstlisting}
\end{scriptsize}

However, in the real-world apps, implementing deep links for activities are much more than declaring an intent filter and adding the logic. Refactoring has to be made on the current implementation of apps. The code snippet in Listing~\ref{code:example} shows an example of a typical Android app. The app has three activities $Main$, $A$, and $B$. $Main$ activity is the entrance of the app. It has two buttons: when $button1$ is clicked, activity $A$ is invoked by an intent with a string parameter $p1$ (intent $i1$); when $button2$ is clicked, a fragment $frag$ is shown on the current $Main$ activity. When activity $A$ is launched, a $fooList$ is initialized with the incoming parameter $p1$. Activity $A$ may invoke activity $B$ with an intent $i2$ to display the detail of a $foo$ when $button3$ is invoked. In the following, we use this example to illustrate the efforts of implementing deep links.

\begin{scriptsize}
\begin{lstlisting}[caption={Motivating Example}, label = {code:example},language=Java]
public class Main extends Activity
                        implements OnClickListener {
   public void onCreate(Bundle savedInstanceState) {
      ...
      button1.setOnClickListener(this);
      button2.setOnClickListener(this);
      ...
   }
   public void onClick(View v) {
      switch (v.getId()) {
      case R.id.button1:
         Intent i1 = new Intent(Main.this, A.class);
         String s1 = ...;
         intent.putExtra("p1", s1);
         startActivity(intent);
         break;
      case R.id.button2:
         Fragment frag = new ChildFragment();
         getFragmentManager().beginTransaction().add(frag);
         break;
      ...
      }
   }
}
public class A extends Activity {
   static List fooList;
   public void onCreate(Bundle savedInstanceState) {
      String s = getIntent().getStringExtra("p1");
      fooList = getFooList(s);
      ...
      button3.setOnClickListener(new OnClickListener() {
         public void onClick(View v) {
            int fooIndex = ...;
            Intent i2 = new Intent(A.this, B.class);
            intent.putExtra("foo", fooIndex);
            startActivity(intent);
         }
      });
   }
}
public class B extends Activity {
   public void onCreate(Bundle savedInstanceState) {
      int fooIndex = getIntent().getIntExtra("foo");
      Foo foo = (Foo) A.fooList.get(fooIndex);
      ...
   }
}
\end{lstlisting}
\end{scriptsize}

$Main$ activity naturally supports deep link because the intent to launch the app always opens the $Main$ activity. For activity $A$, it is relatively easy to add a deep link because the intent to launch $A$ from $Main$ contains only a string parameter $p1$ that can be directly implemented by an implicit intent. However, for activity $B$, although the intent to launch $B$ from $A$ contains only an integer parameter $foo$, the deep link to $B$ cannot be simply implemented by an implicit intent because $B$ relies on the data structure $fooList$ constructed in $A$ to retrieve the specific instance of $Foo$ by $fooIndex$. Such a kind of data dependency is prevalent for Android apps to achieve better performance. As a result, in order to add a deep link to $B$, developers have to refactor the implementation of $B$ to initialize $fooList$, or to add extra logics to retrieve the requested instance of $Foo$ according to $fooIndex$.

Current Android apps usually use fragments to organize the views of an activity. In the motivating example, $Main$ activity has a fragment $frag$ that is shown only when a button is clicked. Adding deep links to only activities is sometimes too coarse-grained. As a result, it is required to reach the specific fragments within an activity by deep links, just like reaching a specific part of a web page by an anchor in the URL. Different from activity transitions based on standard intents, fragment transitions are usually implemented by internal functions that vary among different apps. When adding a deep link to the fragment, developers have to handle the transition to the fragment $frag$, e.g., in a way of encapsulating the fragment transition into a common function and invoking the function according to incoming parameters. Such code refactoring requires re-designing the fragment-management logic of the activity.

On the whole, as we will show later in the empirical study, adding deep links to Android apps requires non-trivial developer efforts. On one hand, some activities cannot be directly reached due to data dependencies on other activities. On the other hand, the prevalence of fragment usage requires fine-grained deep links that can reach specific fragments inside an activity.
\section{Empirical Study of Deep Links}\label{empirical}
\begin{figure*}
  \centering
    \subfigure[Trend of apps with deep links]{
    \label{fig:motivation:trend1} %% label for second subfigure
    \includegraphics[width=0.23\textwidth]{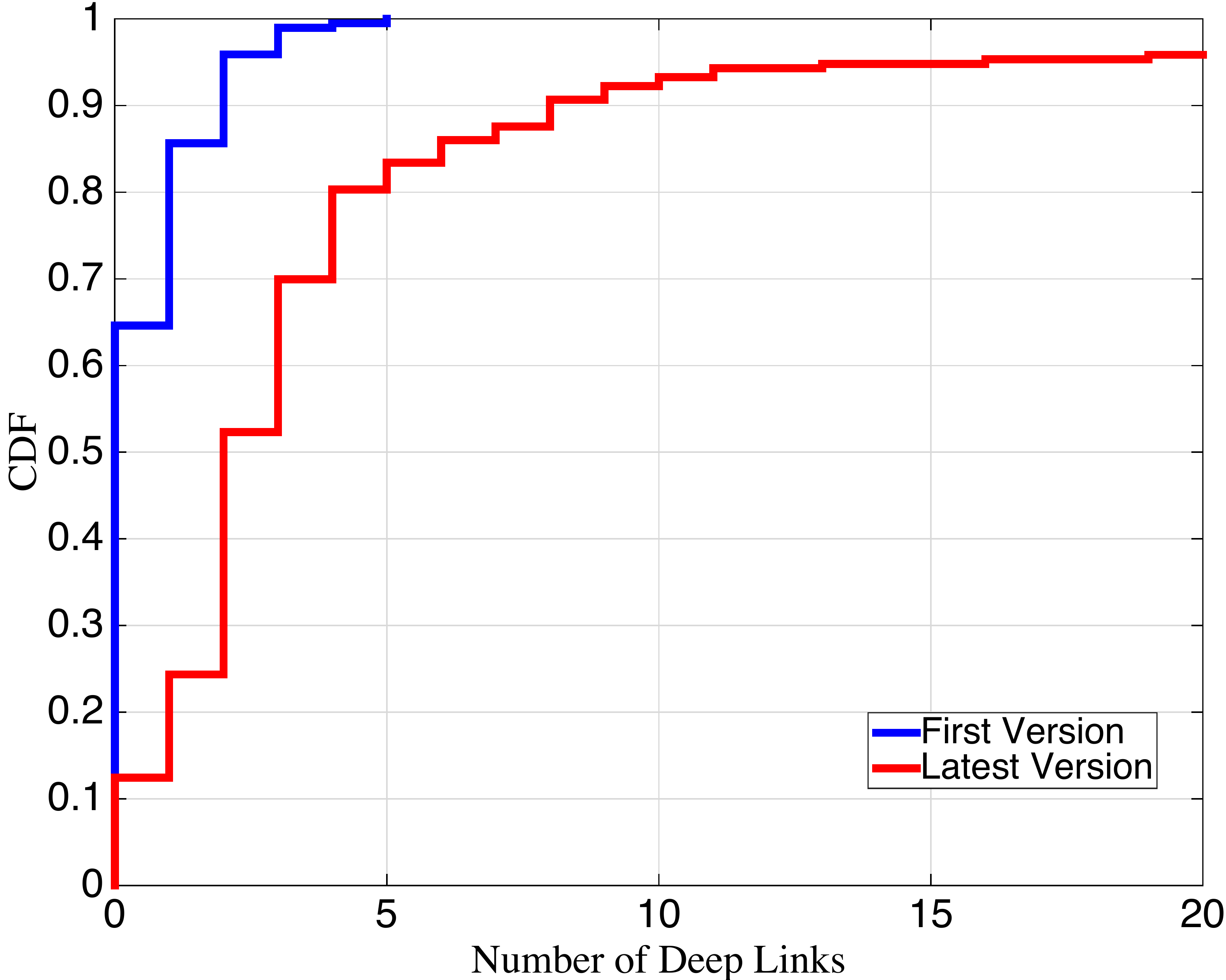}}
    \subfigure[Trend of deep-linked activities in deep-link supported apps]{
    \label{fig:motivation:trend2} %% label for second subfigure
    \includegraphics[width=0.23\textwidth]{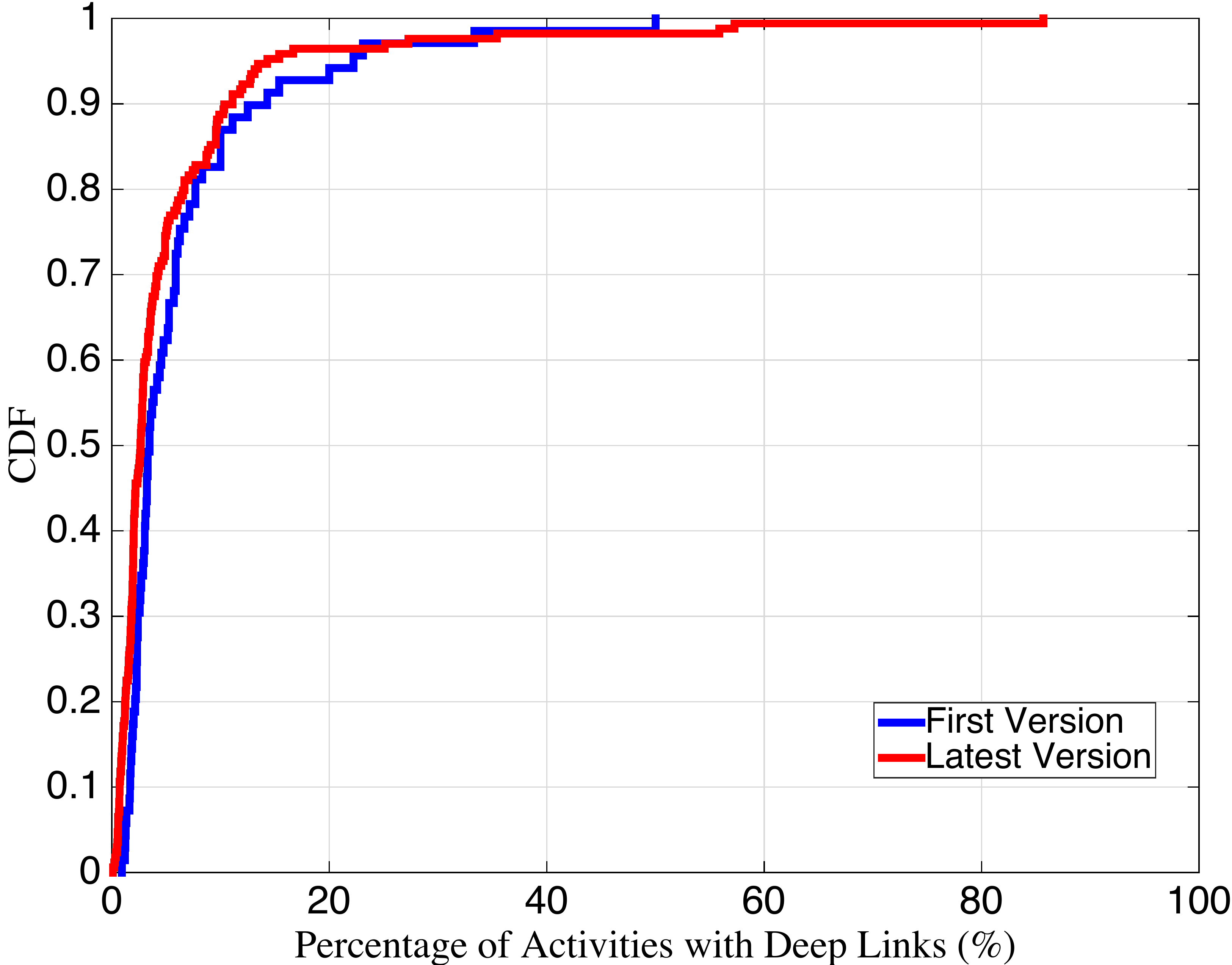}}
    \subfigure[Distribution of the number of deep links]{
    \label{fig:motivation:coverage1} %% label for first subfigure
    \includegraphics[width=0.23\textwidth]{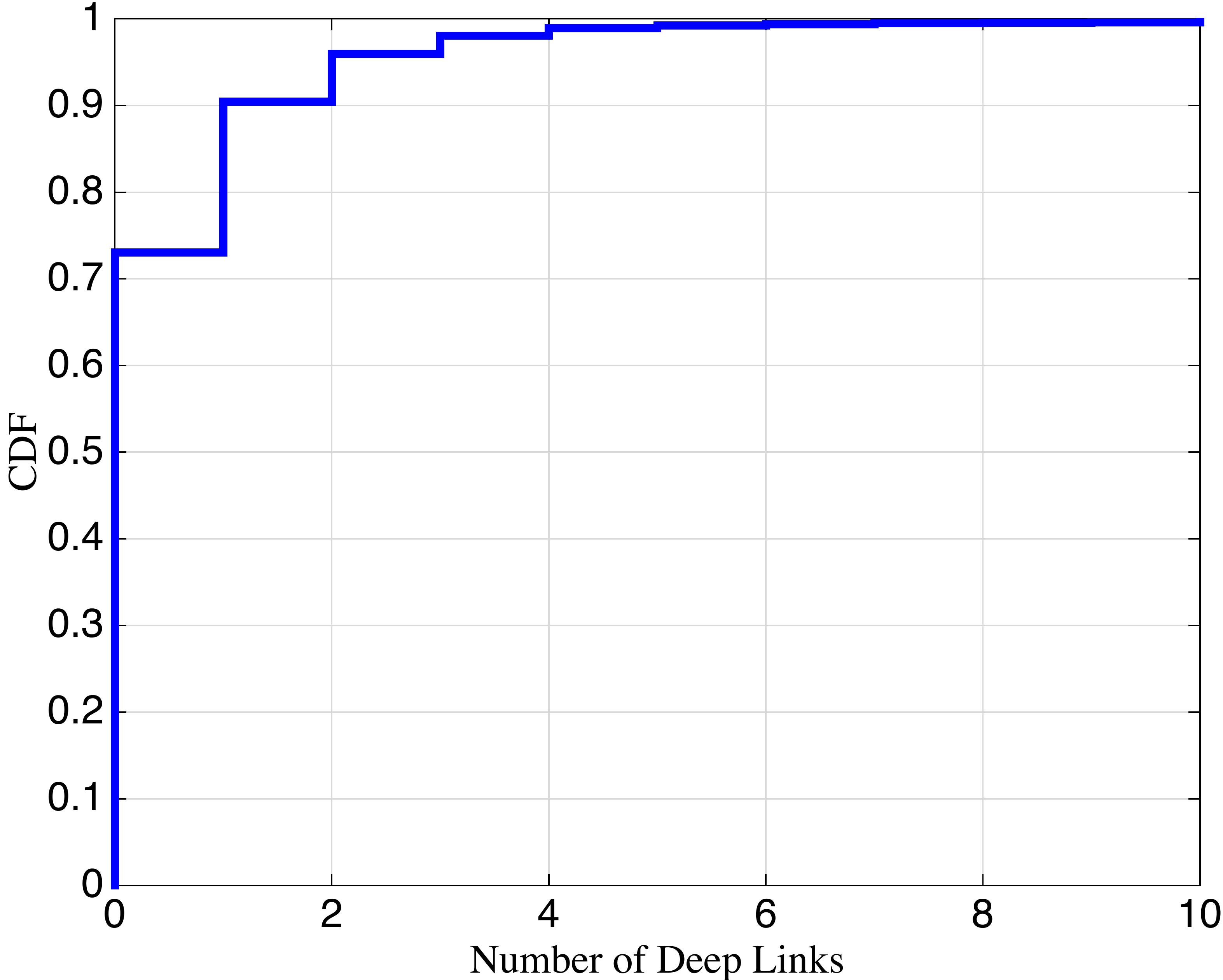}}
    \subfigure[Distribution of the percentage of deep-linked activities]{
    \label{fig:motivation:coverage2} %% label for second subfigure
    \includegraphics[width=0.23\textwidth]{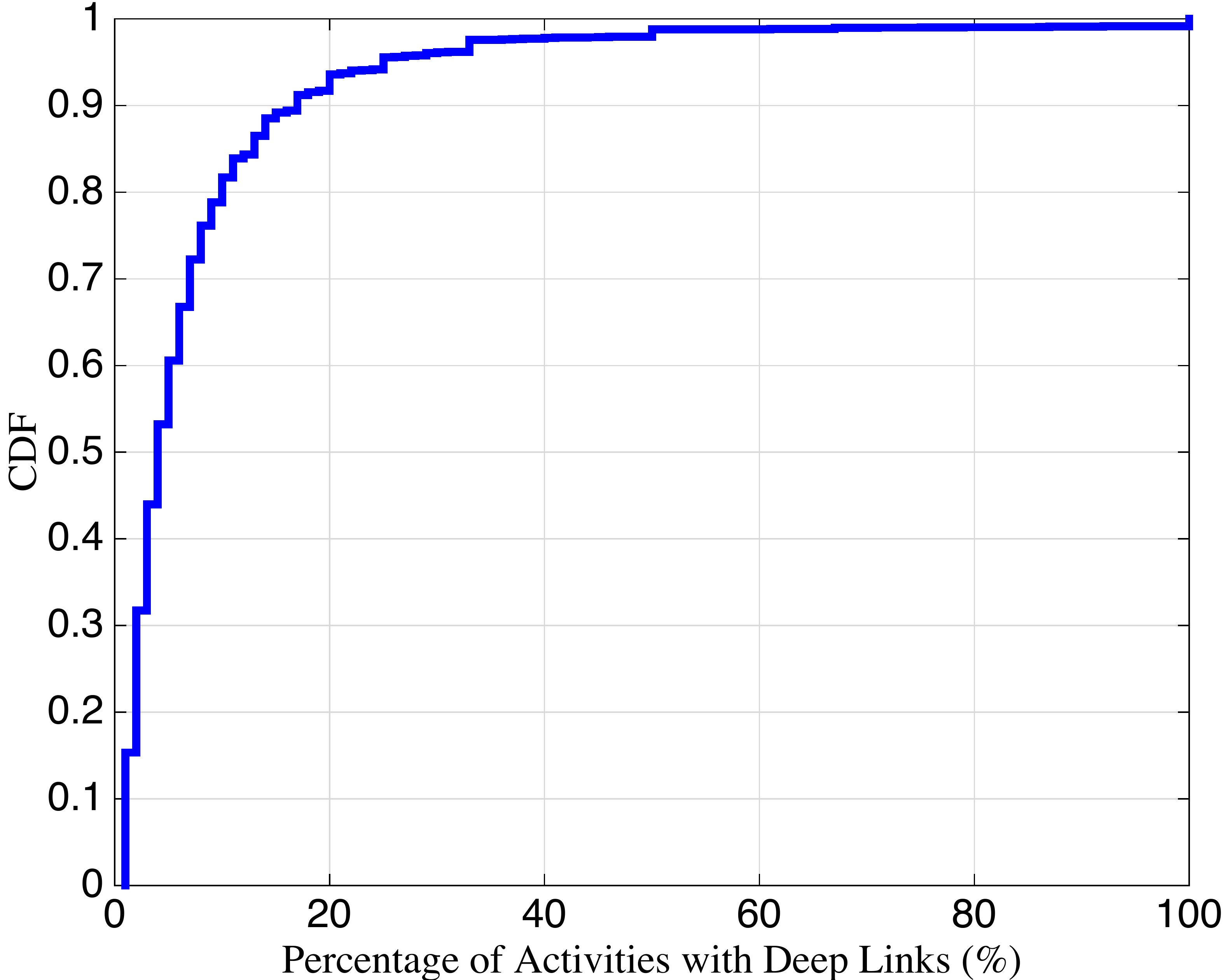}}
  \caption{The trend and status of deep links among Android apps.}
  \label{figure:motivation} %% label for entire figure
\end{figure*}
Given the benefits of deep links for mobile apps, in this section, we present an empirical study to understand the state of practice of deep links in current Android apps. We focus on three aspects: \textit{(1) the trend of deep links with version evolution of apps}; \textit{(2) the number of deep links in popular apps}; (3) \textit{how deep links are realized in current Android apps.}
% and Wandoujia\footnote{Wandoujia is the top Android app store in China, owning more than 300 million users and 1 million free free apps}~\cite{Web:Wandoujialeading}.
% as well as 8 popular open-source apps from GitHub.

%\subsection{Usage Scenarios of Deep Links}
%\begin{figure}[t]
%\centering
%  \includegraphics[width=0.8\columnwidth]{figures/android-app-indexing}
%  \caption{Google App Indexing.}~\label{fig:example}
%\end{figure}

%\textbf{App Indexing}. The goal of app indexing is to put the apps in front of users who use a search engine. When a search result can be served from an app, users who have installed the app can go directly to the page containing the result. Figure~\ref{fig:example} shows an example of Google App Indexing. When a user searches a term ``\textit{triple chocolate therapy recipe}'' on Google, one result comes from the `\textit{`Allthecooks Recipes}'' app. If the user clicks the button ``Open in app'', then the target app page of the search result is directly opened. App Indexing is actually a concrete usage scenario of deep link.

%\textbf{Sharing}. Use Facebook App Links as an example.

%\textbf{Bookmarks}. Use iPhone search box as an example.

%\subsection{Issues of Current Deep Links}
In practice, there is no standard way of measuring how many deep links an app has. However, from Section~\ref{sec:motivation}, we can infer an essential condition, i.e., \textit{\textbf{activities that support deep links MUST have a special kind of intent filters declared in the \texttt{AndroidManifest.xml} file}}. Such a kind of intent filters should use the \texttt{android.intent.}\texttt{action.VIEW} with the \texttt{android.}\texttt{intent.category.BROWSABLE} category. We denote these intent filters as deep-link related. Therefore, we can simply take the number of activities with deep-link related intent filters as an indicator to estimate the number of deep links for an Android app. In other words, if an activity has a deep-link related intent filter, we regard that such an activity has a deep link.
\subsection{Evolution of Deep Links with Versions}
We first validate that the support of deep links is really desired. To this end, we investigate the trend of deep links along with the version evolution. Due to the banned Internet access in mainland China, we cannot perform large-scale study on apps on Google Play. Instead, we choose top 200 apps ranked by Wandoujia, a leading Android app store in China which was studied in our previous work~\cite{WWW15Li,Li:IMC15,Lu:ICSE16,Ubicomp17Lu}. To make comparison, we manually download each app's first version that could be publicly found and its latest version published on its official website as of August 2016. We compare the number of deep links of the two versions of each app. Figure~\ref{fig:motivation:trend1} indicates the change of the number of deep links across these two versions. Generally, it is observed that when the app is first released on the app store, only about 35\% of apps have deep links. In contrast, more than 87\% of these apps have supported deep links in their latest versions. More specifically, the maximum number of deep links is only 5 in the first version of all investigated apps (the app with the most deep links is \emph{Tencent News}). In contrast, 20\% of the investigated apps have more than 5 deep links in their latest versions, and four apps (\emph{Dianping}, \emph{Meituan}, \emph{Taobao}, \emph{Sina Weibo}) have even more than 100 deep links. Such a change indicates the popularity of deep links keeps increasing in the past few years.
\subsection{Coverage of Deep Links}
Although the number of deep-link supported apps increases, the percentage of activities that have deep links in deep-link supported apps is still rather low. We compute the ratio of deep-link supported activities relative to the total number of activities. As shown in Figure~\ref{fig:motivation:trend2}, the distribution of the percentage does not change much between the first version and latest version. Among 90\% of deep-link supported apps, only less than 10\% of activities have deep links. A small difference is that the maximum percentage of deep-linked activities increases from 50\% in the first version to 85\% in the latest version.

Aiming to expand the investigation to a wide scope, we study the latest version of top 20,000 apps ranked by the number of downloads on Wandoujia as of August 2016. Figure~\ref{fig:motivation:coverage1} demonstrates the coverage of deep links among these apps. Similar to the preceding results, about 73\% of the apps do not have deep links, while 18\% of the apps have only one deep link. Such a result indicates that deep links are not well supported in a large scope of current Android apps.

Considering the percentage of deep-linked activities in deep-link supported apps, Figure~\ref{fig:motivation:coverage2} shows that the median percentage is just about 5\%, implying that a very small fraction of the pages can be actually accessed via deep links. Considering the best cases, only 5\% of apps have more than 20\% of their activities support deep links.

In summary, the preceding empirical study demonstrates the low coverage of deep links in current Android apps. Such a result is a bit out of our expectation, since deep links are widely encouraged in industry to facilitate in-app content access. There could be some possible reasons leading to the low coverage of deep links. First, as deep link is a relatively new concept, it may cost some time to be adopted by Android developers. Second, due to commercial or security considerations, developers may not be willing to expose their apps to third parties through deep links. Third, the developers do not have clear motivation to determine which part of their apps needs to be exposed by deep links. Fourth, as we will show later, implementing deep links requires non-trivial developer effort so that developers may not be active to introduce deep links in their apps. However, deep link is still promising in the mobile ecosystem given the strong advocation by major Internet companies as well as the potential revenue brought by opening data and cooperating with other apps.

\subsection{Developer Effort}
Indeed, supporting deep links requires the developers to write code and implement the processing logics. In practice, there are usually two ways of implementing deep links. One is to establish implicit intents for each activity that needs to be equipped with deep links. The other is to use a central activity to handle all the deep links and dispatch each deep link to its target activity. Although there have already been some SDKs for deep links~\cite{mobiledeeplinking,DeepLinkDispatch}, implementing deep links exactly requires the modifications or even refactoring of the original implementation of the apps. We then investigate the actual developer effort when releasing deep links for an Android app.

For simplicity, we study the code evolution history of open-source apps on GitHub. We search on GitHub with the key word ``\textit{deep link}'' among all the code-merging requests in the Java language. There are totally 4,514 results returned. After manually examining all the results, we find 8 projects that actually add deep links in their code-commit history. We carefully check the code changes in the commit related to deep links. Table~\ref{table:locgithub} shows the LoC (lines of code) of changes in each project when adding one deep link in the corresponding code commit. The changes include the addition, modification, and deletion of the LoC.  We can observe that the least number of the LoC is 45 and the biggest can reach 411. For the app \texttt{SafetyPongAndroid}\footnote{\url{https://github.com/SafetyMarcus/SafetyPongAndroid}}, we find that a large number of changes attribute to the refactoring of app logics to enable an activity to be directly launched without relying on other activities. Such an observation can provide us the preliminary findings that developers need to invest non-trivial manual efforts on existing apps to support deep links. Such a factor could be one of the potential reasons why deep links are of low coverage.
\begin{table}[t]
\centering
\caption{LoC changes when adding deep links of open-source apps on GitHub.}\label{table:locgithub}
\begin{tabular}[t]{l|r}
 \hline
  \textbf{Repository Name} & \textbf{LoC Changes}\\
  \hline
  stm-sdk-android & 45 \\
  mobiledeeplinking-android & 62 \\
  WordPress-Android & 73 \\
  mopub-android-sdk & 78 \\
  ello-android & 87 \\
  bachamada & 179 \\
  sakai & 237 \\
  SafetyPongAndroid & 411 \\
  \hline
\end{tabular}
\end{table}
%\subsection{Implications from the Study}
%Current support of deep links is rather limited. There are two issues that should be resolved.

\section{Aladdin: In a Nutshell}\label{sec:approach}
The findings of our empirical study demonstrate the \textbf{low coverage} and \textbf{non-trivial developer efforts} of supporting deep links in current Android apps. To improve coverage, one possible solution is to  leverage the program analysis of apps to extract how to reach locations of in-app contents pointed by deep links. However, the static analysis of Android apps can only build structure relations among activities but cannot analyze dynamic fragments inside an activity; the dynamic analysis of Android apps can analyze dynamic fragments but suffer from low coverage of activities, i.e., only a small fraction of activities can be reached by dynamic analysis.

\begin{figure*}[t]
\centering
  \includegraphics[width=1\textwidth]{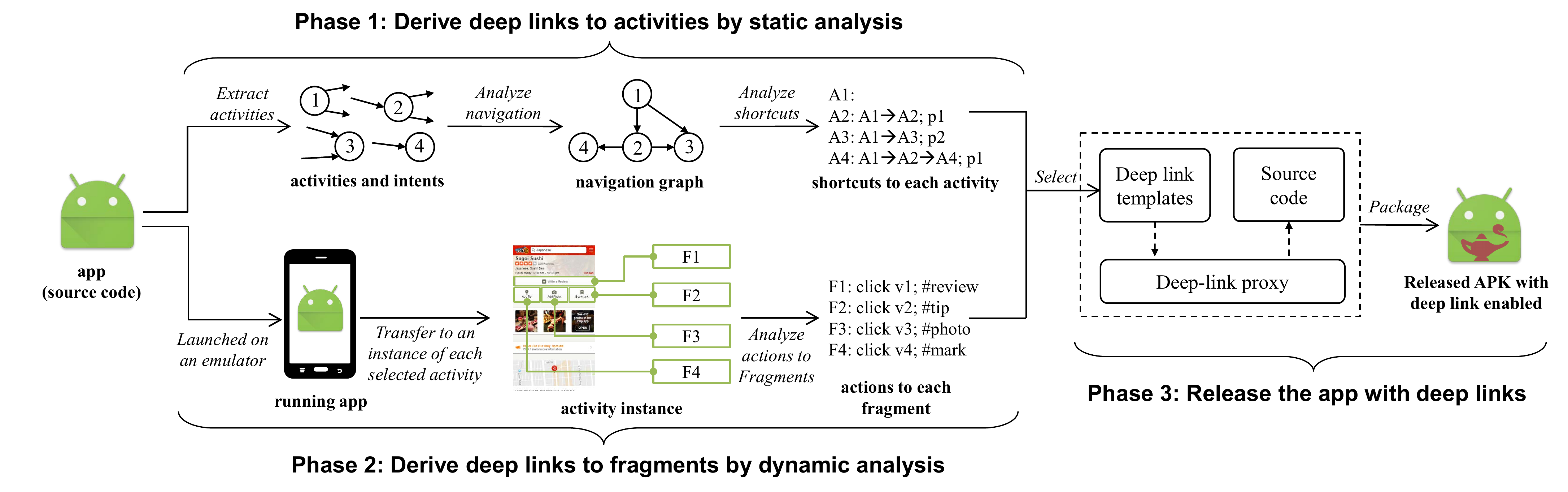}
  \caption{Approach Overview.}~\label{fig:approach}
\end{figure*}

To address the challenge, we propose a cooperative framework and design a tool \textbf{\emph{Aladdin}} to automate the release of deep links. Our cooperative framework combines static analysis and dynamic analysis while minimally
engaging developers to provide inputs to the framework for automation. Figure~\ref{fig:approach} shows the overview of our approach, and the workflow is illustrated as follows.

Given the source code of an app, Aladdin first derives deep link templates that record the scripts of how to reach arbitrary locations inside the app. Each template is associated with an activity and consists of two parts: one part is the intent sequence extracted based on static analysis, recording how to reach the activity from the main activity of the app; the other part is the action sequence extracted based on dynamic analysis, recording how to reach fragments in the activity from the entrance of the activity. After developers configure to verify the activities and fragments to be deep linked, the corresponding templates and a deep-link proxy are automatically generated, and are packaged with the original source code of the app as a new \texttt{.apk} file. Each template has a URI schema that is used to populate a concrete deep link. The deep-link proxy provides a replay engine that can replay the sequences at runtime to execute the deep links.

When a deep link is requested with a URI conforming to a schema, the deep-link proxy is triggered to work. The corresponding template is instantiated by assigning values to parameters in the template. Then the proxy first makes the app transfer to the target activity by issuing the intents one by one in the intent sequence. Then the proxy makes the activity transfer to the target fragment by performing the actions one by one in the action sequence. Finally, the target location could be reached. Such a proxy-based architecture does not modify any original business logic of the app and is coding-free for developers.

We next present the details of each phase.
\subsection{Deriving Deep Links to Activities}\label{sec:activity}
Essentially, reaching an activity with a deep link is to issue one intent that could launch the target activity. However, the complexity of activity communications in Android apps makes it not easy to use a single intent to reach an activity for existing apps. For example, a target activity may rely on internal objects that are created by previous activities in the path from the main activity to the target activity. To address the issue, rather than the single intent to an activity, Aladdin abstracts a deep link to an activity as a sequential path of activity transitions via intents, starting from the main activity of the app to the target activity pointed by the deep link. Therefore, the activity dependency can be resolved because we actually follow the original logic of activity communications.

\subsubsection{Navigation Analysis}

Since activities are loosely coupled that are communicated through intents, there is no explicit control flows between activities. Therefore, we design a \textit{Navigation Graph} to abstract the activity transitions. Here we give the formal definitions of activity transition and navigation graph.
\noindent \begin{Definition}[Activity Transition]
An activity transition $t(\mathcal{L})$ is triggered by an intent, where $\mathcal{L}$ is the combination of all the fields of the intent including action, category, data, and objects of basic types from the extra field.
\end{Definition}

Since an intent essentially encapsulates several messages passed between two activities, we use a label set $\mathcal{L}$ to abstract an intent. Two intents are equivalent if and only if the label sets are completely the same. Note that, from the extra field which is the major place to encapsulate messages, we take into account only the objects of basic types including \texttt{int}, \texttt{double}, \texttt{String}, etc. The reason is that objects of app-specific classes are usually internally created but cannot be populated from outside of the app. As a result, this kind of intents cannot be replayed at runtime.
\begin{Definition}[Navigation Graph]
A Navigation Graph $G$ is a directed graph with a start vertex. It is denoted as a 3-tuple, $G<V,E,r>$, where $V$ is the set of vertices, representing all the activities of an app; $E$ is the set of directed edges, and every single $e(v_1,v_2)$ represents an activity transition $t(\mathcal{L})$; $r$ is the start vertex.
\end{Definition}

In such a navigation graph, the start vertex $r$ refers to the main activity of the app. We assume that each node in $V$ could be reachable from the start vertex $r$. The navigation graph can have multi-edges, i.e., $\exists e_1,e_2\in E, v_{start}(e_1)=v_{start}(e_2)~and~v_{end}(e_1)=v_{end}(e_2)$, indicating that there can be more than one transition between two activities. In addition, it should be noted that the navigation graph can be cyclic.

\subsubsection{Shortcut Analysis}
After constructing the navigation graph, we can analyze the paths to each activity.
\begin{Definition}[Path]
A path to an activity $\mathcal{P}_a$ is an ordered list of activity transitions $\{t_1, t_2, \dots, t_k\}$ starting from the main activity, where $k$ is the length of the path.
\end{Definition}

According to the path definition, the activity transition $t_1$ is always the app-launching intent that opens the main activity. The path $\mathcal{P}_a$ can ensure that all the internal dependencies are properly initialized before reaching the activity $a$.

\begin{figure}[t]
\centering
  \includegraphics[width=0.5\columnwidth]{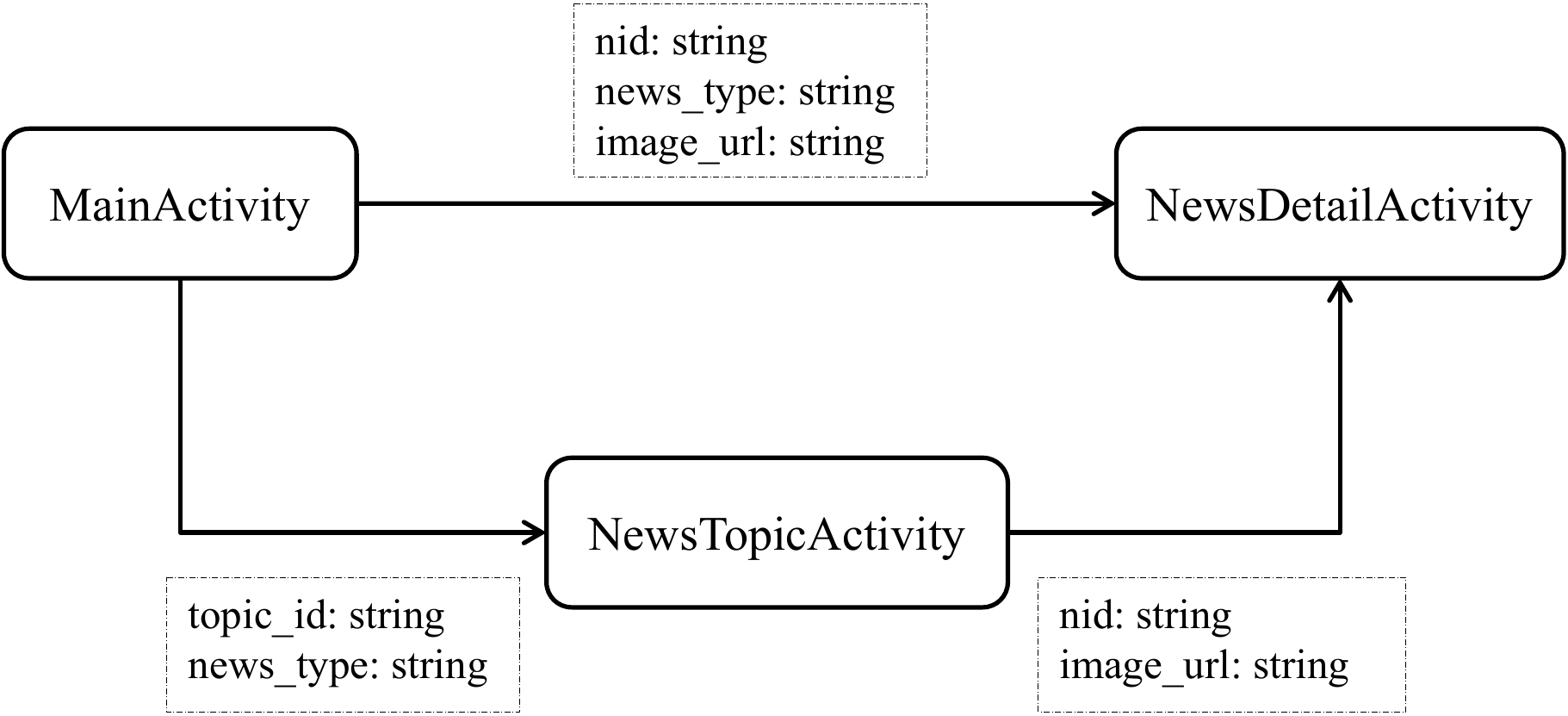}
  \caption{Example of Wallstreet News App.}~\label{fig:wallstreet}
\end{figure}

In practice, there can be various paths to reach a specific activity. Figure~\ref{fig:wallstreet} shows the example in the ``\textit{Wallstreet News}'' app. If we want to reach a news page (\texttt{NewsDetailActivity}), there are two paths. One is to navigate directly from the \texttt{MainActivity}. The other is to switch to the topic page of \texttt{NewsTopicActivity} and then navigate to the news page. Obviously, the former path is shorter than the latter one. Since our approach uses the activity transitions to reach activities by deep links, the path should be as short as possible to reduce the execution time at system level. However, in some circumstances, a shorter path to an activity cannot cover all the instances of the activity. For example, the intermediate activities on the path can depend on some internal variables, so that they can be reached only via a longer path where these variables are assigned.

We define that the path $p_1$ can replace the path $p_2$ if and only if $\mathcal{L}_{p_1}\subset \mathcal{L}_{p_2}$. Here $\mathcal{L}_{p_j}$ is the combination of all the labels in the path $p_j$: $\mathcal{L}_{p_j}=\cup\{\mathcal{L}(t_i)|t_i\in p_j\}$. For example, in the ``\textit{Wallstreet News}'' app, the path 1 from the \texttt{MainActivity} to the \texttt{NewsDetailActivity} has three labels, i.e., ``\textit{nid}'', ``\textit{image\_url}'', and ``\textit{news\_type}''. We can observe that all these three labels are included in path 2 which passes the \texttt{NewsTopicActivity}. Hence, we can use path 1 to replace path 2.

With the definition of path replacement, we can define the possible shortest path as a Shortcut. We can use the shortcut to replace the path to an activity accounting for shorter execution time.
\begin{Definition}[Shortcut]
A Shortcut of a path $\mathcal{T}(p)$ is the shortest path that can replace path $p$.
\end{Definition}

We design the Algorithm~\ref{algo:shortcut} for finding the shortcuts in navigation graph $G$. For every vertex in the graph, we get its paths (Line 2) and sort the paths by their length in ascending order (Line 3). Then we enumerate every path in the path list, to find the shortest path that can replace it (Line 4-12), if they fulfill the requirement of label containing (Line 7). Due to the increasing sequence, the first available path that we obtain is the shortest one. We store it in a two-dimension map structure (Line 8).
\begin{algorithm}
\small
    \caption{Shortcut computation.}
    \label{algo:shortcut}
    \SetAlgoLined
    \KwIn{Navigation Graph $G<V,E,r>$ }
    \KwOut{Shortcut $\mathcal{C}$}
        \ForEach{$v\in V$} {
            $Plist\leftarrow G.path(v)$;\\
            sort\_by\_length($Plist$);\\
            \For{$i\leftarrow 1~to~Plist.length$} {
                $Shorcut[<v, p_i>]\leftarrow p_i$;\\
                \For{$j\leftarrow 1~to~i-1$} {
                    \If{$\mathcal{L}_{p_j}\subset\mathcal{L}_{p_i}$} {
                        $Shorcut[<v, p_i>]\leftarrow p_j$;\\
                        break;
                    }
                }
            }
        }
\end{algorithm}

After computing the shortcuts to an activity, we filter out the unique shortcuts and then put all of them into the intent sequence in the deep link template corresponding to the activity. The labels in the intents can be configured by developers as parameters. At runtime, these parameters are assigned values to reconstruct the shortcut.

\subsection{Deriving Deep Links to Fragments}\label{sec:fragment}
As shown in Section~\ref{sec:background}, there are different fragments in an activity served as user interface, just like the frames in a web page. In order to reach a specific fragment directly with a deep link, we should further analyze how to transfer to fragments of an activity.

Contrary to activity transitions where intents can be sent to invoke the transition, the fragment transitions often occur after users perform an action on the interface such as clicking a view, then the app gets the user action and executes the transition. Due to the dynamics of activities, fragments of activities may be dynamically generated, just like AJAX on the Web. To the best of our knowledge, it is currently not possible to find out fragments by static analysis. Thus, we tend to use dynamic analysis, traversing the activity dynamically by clicking all the views on the page in order to identify all the fragments and their corresponding trigger actions.

%After developers select the activity to implement deep links, a simulator will be launched and build the graph. Aladdin will list all the fragments and paths to developers for them to choose.

\subsubsection{Fragment Identification}

Unlike activities where className is the identifier of different activities, fragments usually do not have explicit identifiers. To determine whether we have switched the fragment after clicking a view, we use the view structure to identify a certain fragment. In Android, all the views are organized in a hierarchy view tree. We get the view tree at runtime and design Algorithm~\ref{algo:structhash} to calculate the structure hash of this tree, and use the hash to identify the fragments. The algorithm is recursive with a view $r$ as input. If $r$ does not have children, the result is only the string hash of $r$'s view tag (Line 2). If $r$ has children (Line 3), then we use the algorithm to calculate all the hash of its children recursively (Lines 5-7). Then, we sort the $r$'s children based on their hash values to ensure the consistency of the structure hash, because a view's children do not keep the same order every time(Line 8). Next, we add each children's hash together with the view tag forming a new string (Line 10), and finally return the string hash (Line 13). When inputing the root view of the tree to the algorithm, we could get a structure hash of the view tree. The hash can be used as an identifier of a fragment.

\begin{algorithm}
\small
    \caption{Computing structure hash of view tree.}
    \label{algo:structhash}
    \SetAlgoLined
    \KwIn{View $r$}
    \KwOut{Structure Hash $h$}
        \textbf{function}~TreeHash($r$)\\
        $str \leftarrow r.viewTag$\\
        \If{$r.hashChildren()$} {
            $children \leftarrow r.getChildren()$\\
            \ForEach{$c \in children$} {
                $c.hash \leftarrow TreeHash(c)$
            }
            $children \leftarrow sort\_by\_hash(r.getChildren())$\\
            \ForEach{$c \in children$} {
                $str += c.hash$
            }
        }
    \Return{$hash(str)$}
\end{algorithm}

\subsubsection{Fragment Transition Graph}
In order to retrieve all the fragments as well as triggering actions to each fragment, we define a fragment transition graph to represent how fragments are switched in an activity.
\begin{Definition}[Fragment Transition Graph]
A Fragment Transition Graph is a directed graph with a start vertex. It is denoted by a 3-tuple, $FTG <V, E ,r>$. $V$ is the set of vertices, representing all the fragments of an activity, identified by the structure hash. $E$ is the set of directed edges. Each edge e is a 3-tuple, $e <S, T, I>$. $S$ and $T$ are source and target fragments where $I$ represents the resource id of the view which invokes the transition. $r$ is the start vertex.
\end{Definition}

The dynamic analysis performs on an instance of an activity. Therefore, after developers select the activity to support deep links to fragments, a simulator will be launched and developers are asked to transfer to an instance page of this activity. From this page, the simulator traverses the activity in the depth-first sequence as the algorithm~\ref{algo:FTGGeneration} describes. For each view in the current fragment, we try to click it (Lines 2-3) and check whether the current state has changed. If the activity has changed, then we can use the system function \texttt{doback()} to directly return to previous condition (Line 5). Otherwise, we check the fragment state. If the structure hash of the current fragment is different from that of the previous one, the fragment has changed (Lines 8-9). Therefore, we can add it into the edge set if it is a new fragment (Line 10-12). The dynamic analysis is similar to the web crawlers that need to traverse every web page, except that Android only provides a \texttt{doback()} method that can return to the previous activity, but not to the previous fragment. So to implement backtrace after fragment transitions, we have to restart the app and recover to the previous fragment (Line 14).

\begin{algorithm}
\small
    \caption{Generation of the Fragment Transition Graph.}
    \label{algo:FTGGeneration}
    \SetAlgoLined
    \KwIn{Activity $a$, Fragment $f$}
    \KwOut{Fragment Transition Graph $G$}
        \textbf{function}~FTGBuild($a, f$)\\
        \ForEach{$v\in f.views()$} {
            $v.click()$\\
            \If{$getCurrentActivity()\neq a$} {
                $doback()$\\
                \textbf{continue}
            }
            $cf\leftarrow getCurrentFragment()$\\
            \If{TreeHash($cf.root$)$\neq$TreeHash($f.root$)} {
                \If{$cf \notin G.V$} {
                    $G.E.add(<f, cf, v>)$\\
                    $FTGBuild(a, cf)$
                }
                $recover()$\\
               \textbf{continue}
            }
        }
\end{algorithm}

After finishing the traverse search, we can get the fragment transition graph and a list of fragments. To get the path towards a certain fragment, we simply combine all the edges from the start vertex to the fragment. Thus, developers can choose any fragment to form a deep link by putting the actions into the action sequence in the deep link template.

\subsection{Releasing Deep-Link Supported Apps}\label{sec:packager}
After computing the shortcuts to activities and actions to fragments, the next step is to create the target \texttt{.apk} that supports processing deep links at runtime. Note that developers may want to create deep links to only some locations of their apps. Therefore, Aladdin allows developers to check which ones need the support of deep links.

Then Aladdin generates an abstract URI of the deep links for each selected activity. To be discovered by Android system, the URI should follow the format of ``\textit{scheme://host/path}'' where \emph{scheme}, \emph{host} and \emph{path} could be any string value. In order to conform to the latest app links specification of Android 6~\cite{applinks}, we employ the format of ``\textit{http://host/target?} \textit{parameter\#fragment}'' as the schema of the abstract URI. We use the reverse string of the \texttt{packageName} (usually the domain of the corresponding website) for the \emph{host} field and the \texttt{className} of the activity for the \emph{target} field. All the instances of an activity share the similar prefix before ``?'' but are different on the \emph{parameter} part. For deep links to fragments, the name of the target fragment is after a $\#$. With the abstract URI, we can generate intent filters in AndroidManifest.xml file to handle the corresponding deep links.
\begin{figure}[t]
\centering
  \includegraphics[width=0.6\textwidth]{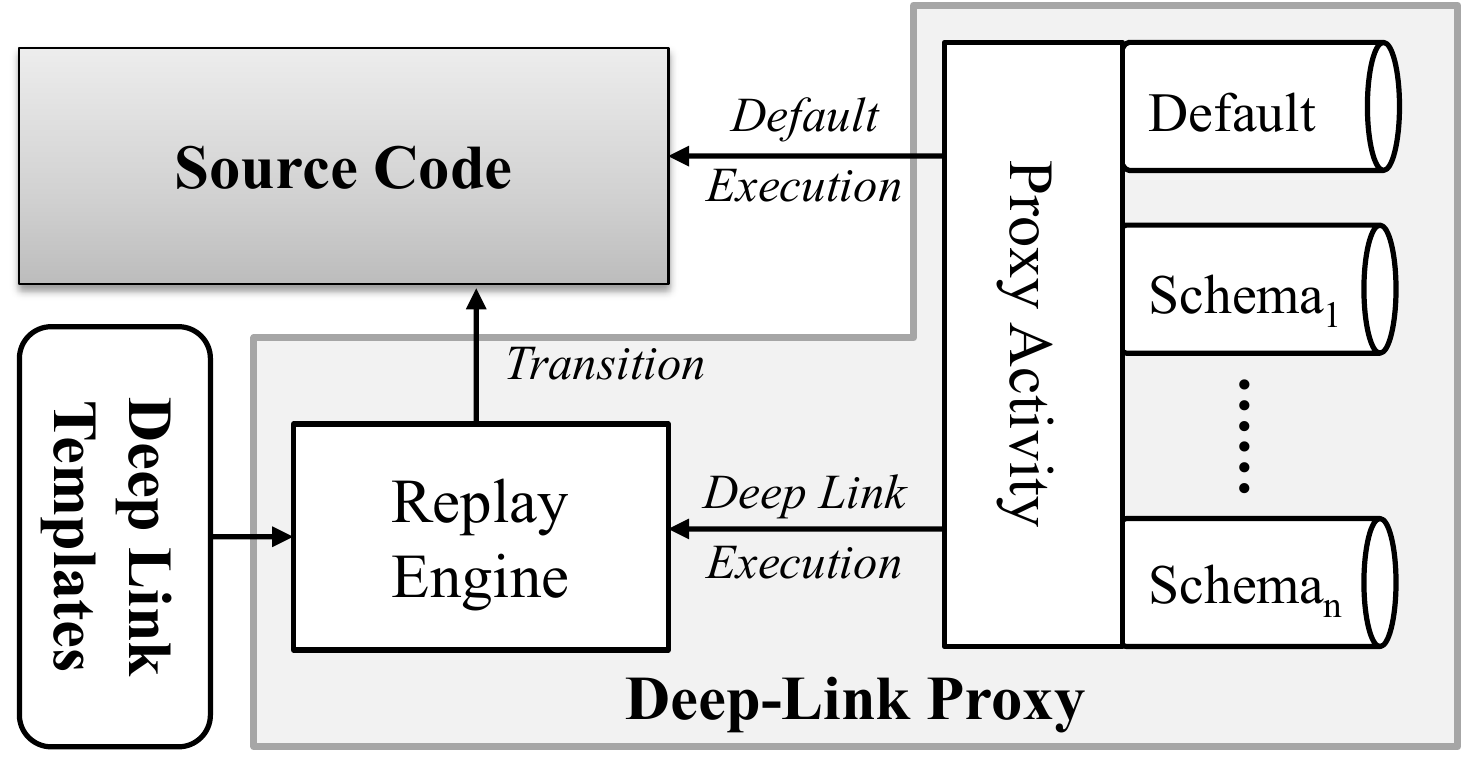}
  \caption{Structure of the app with deep link enabled.}~\label{fig:package}
\end{figure}

Figure~\ref{fig:package} depicts the structure of the created \texttt{.apk}. We leverage a proxy architecture to realize minimal refactorings to the original app. A \textit{Proxy Activity} is used to handle all the incoming requests. The \emph{Proxy Activity} is configured to intent filters that conform to the URI schemas. When an intent is passed to the \textit{Proxy Activity}, if the intent matches one of the schemas, the \textit{Proxy Activity} informs the \textit{Replay Engine} to execute the deep link. If the incoming intent cannot match to any of the schemas, it is then forwarded directly to the original \textit{Source Code} for default execution.

\begin{figure}[t]
\centering
  \includegraphics[width=0.6\textwidth]{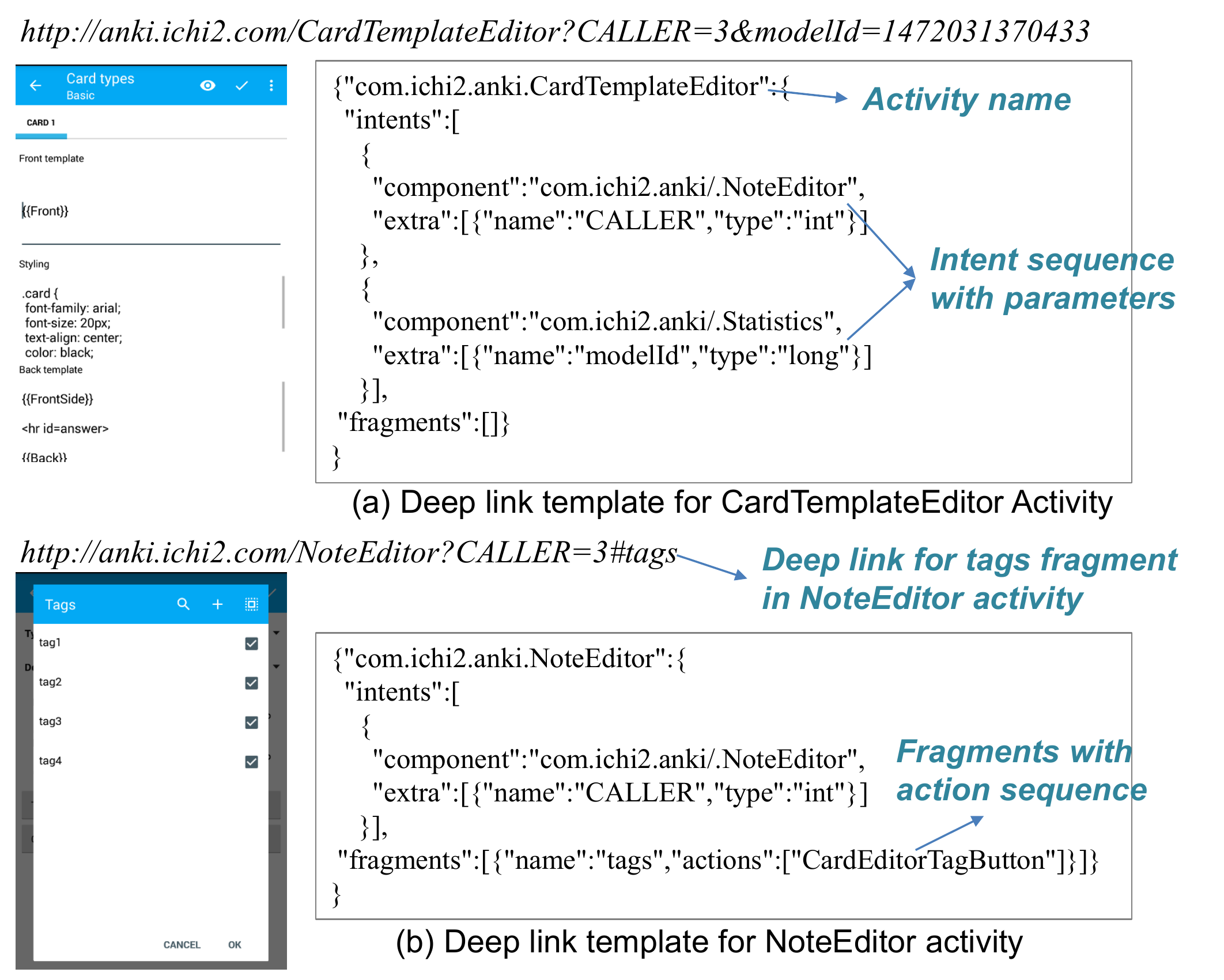}
  \caption{Example of deep link templates and concrete deep links.}~\label{fig:dpexample}
\end{figure}

Figure~\ref{fig:dpexample} shows two deep link templates of the \texttt{Anki} app released by Aladdin. The two pieces of code in the box are deep link templates for \texttt{CardTemplateEditor} and \texttt{NoteEditor} activity. Two intents with two parameters \texttt{CALLER} and \texttt{modeId} have to be issued before reaching \texttt{CardTemplateEditor}. Therefore, the deep link to \texttt{CardTemplateEditor} should explicitly specify the values of the two parameters. Then Aladdin can populate the proper intents to transfer to the target activity. \texttt{NoteEditor} has a fragment naming ``\emph{tags}''. The actions to the fragment is clicking the view whose resource id is \texttt{CardEditorTagButton}. Therefore, to reach the tags fragment, not only should the value of intents be assigned ($CALLER=3$), but the fragment should be specified as well ($\#tags$).

\subsection{Executing Deep Links by Replay at Runtime}
When the deep link is executed, the corresponding deep-link template is instantiated with concrete values to create a replay script. Then the \textit{Replay Engine} communicates with the original \textit{Source Code} and instructs the app to transit through activities and perform actions on views according to the script.

For example, in Figure~\ref{fig:dpexample}(b), when the deep link \url{http://anki.ichi2.com/NoteEditor?CALLER=3#tags} is requested, it implies that the user may want to reach the tags fragment of \textit{NoteEditor} in \texttt{Anki app}. So the Replay Engine first issues the intent with the parameter \texttt{CALLER} is 3. Then the Replay Engine performs a click on the view whose resource id is \texttt{CardEditorTagButton}. Finally, the user can reach the target location.

\section{Implementation}\label{sec:implementation}
In this section, we present the implementation details of Aladdin.

To construct the \textit{Navigation Graph}, we first use the \texttt{IC3} tool~\cite{IC3:ICSE2016} to extract all the activities and intents from the source code of an Android app. Then we apply the \texttt{PRIMO} tool~\cite{primo:POPL2016} to compute the links among activities. For each link computed by \texttt{PRIMO}, we add the corresponding activity as a node to the \textit{Navigation Graph} and connects the two nodes with an edge. The labels on the edge are retrieved from the output of \texttt{IC3}. In particular, some edges have only a sink activity, indicating that this activity can be directly opened from outside of the app. For these edges, we add an edge from the main activity node, to make all the nodes reachable from the main activity.

We use the instrumentation test framework provided by Android SDK to implement the dynamic analysis. Android instrumentation can load both a test package and the App-Under-Test (AUT) into the same process. With this framework, we are able to inspect and change the runtime of an app, such as retrieving view components of an activity and triggering a user action on the target app.

To generate the \texttt{.apk} with deep link enabled, the \textit{Replay Engine} is actually implemented as a test case of the instrumentation test to execute the scripts. The deep-link proxy is implemented as a normal Android activity and the corresponding schemas are regularly configured as intent-filters in the \texttt{AndroidManifest.xml} file. When the proxy activity receives a deep link request, it launches the instrumentation test to run the Replay Engine. The Replay Engine uses the instrumentation object provided by the framework to send intents in the app process, reaching the target activity. Then it sends action events to finally reach the target fragment.

\begin{figure}[t]
\centering
  \includegraphics[width=0.7\textwidth]{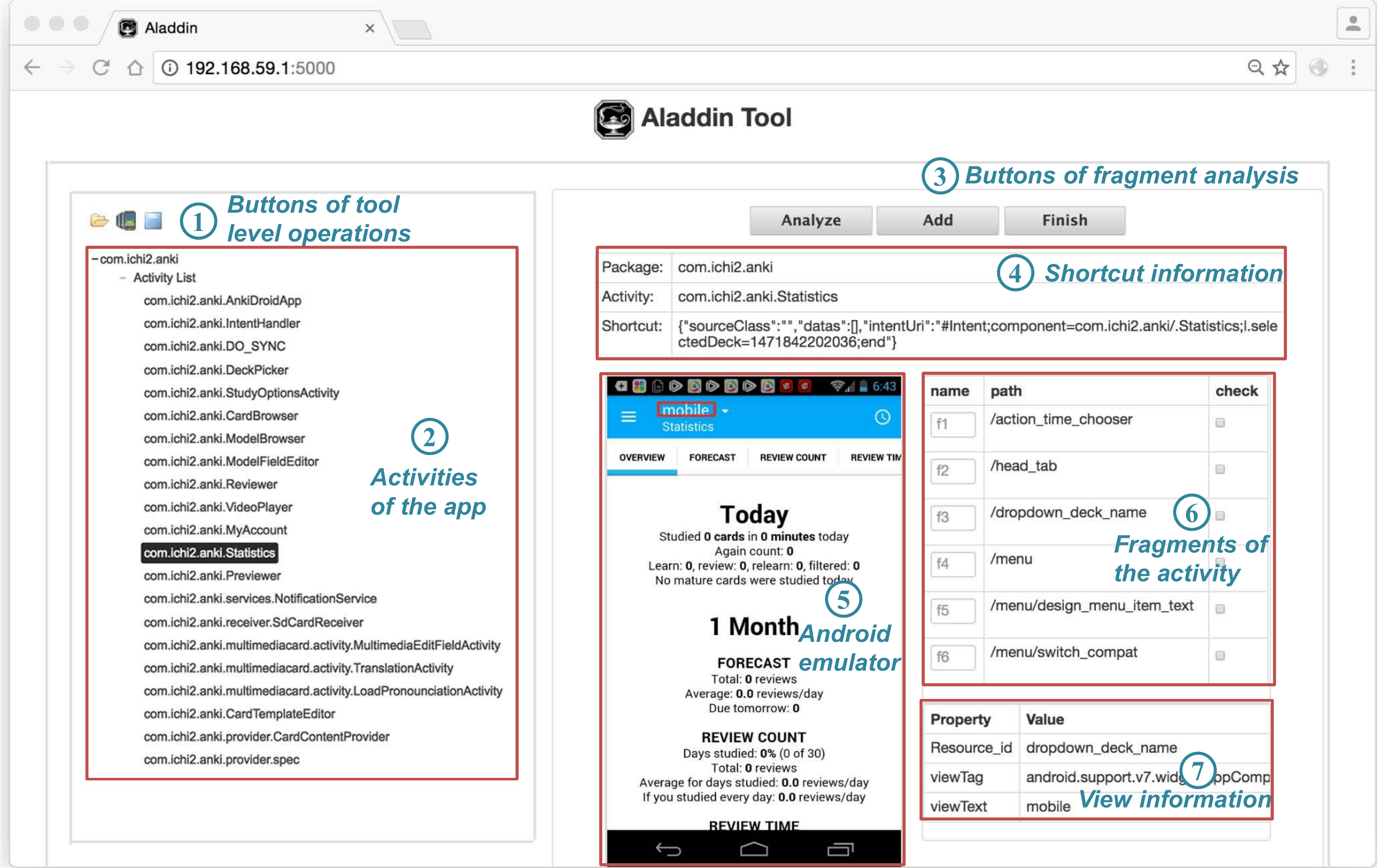}
  \caption{Snapshot of Aladdin tool.}~\label{fig:tool}
\end{figure}

Figure~\ref{fig:tool} shows the snapshot of Aladdin tool. Developers can choose the source code of apps to be analyzed from the button in \textcircled{\scriptsize{1}}. Then the static analysis is automatically performed to extract all the activities (shown in \textcircled{\scriptsize{2}}) as well as the shortcuts of each activity. When an activity is selected, the shortcut information is shown in \textcircled{\scriptsize{4}}. If a deep link to fragments is required, Aladdin allows developers to launch an emulator from \textcircled{\scriptsize{1}}, then transfer to an instance of the activity by performing actions in \textcircled{\scriptsize{5}}. Next, if the dynamic analysis is chosen to run from \textcircled{\scriptsize{3}}, the executed results are listed in \textcircled{\scriptsize{6}}. Developers can explore whether each fragment is correct by clicking each result in \textcircled{\scriptsize{5}}. Developers can also manually add a fragment by performing the actions in \textcircled{\scriptsize{5}} with view information in \textcircled{\scriptsize{7}}. After selecting all the fragments that need deep links, developers click finish button from \textcircled{\scriptsize{3}} to complete the analysis for an activity. When the deep links for activities have been confirmed, developers can click the ``Save'' button in \textcircled{\scriptsize{1}} and package the \texttt{.apk} file as output.

\section{Evaluations}\label{sec:evaluation}
In this section, we evaluate the feasibility and efficiency of Aladdin. First, we investigate how high coverage of activities  Aladdin can achieve to automatically expose deep links based on static analysis. Then, we examine the accuracy of Aladdin's dynamic analysis of fragments in order to generate fine-grained deep links to fragments. Finally, we measure the runtime overhead incurred by Aladdin.

%Our evaluations address the following research questions:
%\begin{itemize}
%\item {\textbf{RQ1}: How high coverage of activities can Aladdin achieve?}
%\item {\textbf{RQ2}: How accurate is Aladdin's dynamic analysis of fragments?}
%\item {\textbf{RQ3}: How high runtime overhead does Aladdin's proxy activity incur?}
%\end{itemize}

\subsection{The App Dataset}
We evaluate Aladdin on top apps chosen from Google Play. Aladdin is applied on the source code of an Android app. However, most of top apps on Google Play are not open source. Although there are some disassembly tools that can generate Java-like code from the \texttt{.apk} file, many apps still fail to be analyzed by Aladdin due to obfuscation or proguard. Note  that the limitation of applying Aladdin on a closed source app is not due to the complexity of the app, but the recurrence of bytecode obfuscation in apps. In order to make the evaluations comprehensive, we choose one popular app from each category of Google Play except games. We prefer open source apps with the number of downloads exceeding 100K. If no open source apps can meet the requirement, then we choose alternatives that can be successfully analyzed by Aladdin. Table~\ref{table:apps} shows all the apps chosen for evaluations. There are 20 apps in total for evaluations\footnote{We are also applying Aladdin to popular open-source apps from F-droid. Please refer to \url{http://sei.pku.edu.cn/~mayun11/aladdin/} for more details.}.

\begin{table*}[t]\tiny
\caption{Selected apps for evaluations.}\label{table:apps}
\begin{tabular}[t]{llllrrrr}
 \hline
  \textbf{App} & \textbf{Category} & \textbf{Description} & \textbf{Downloads} & \tabincell{c}{\textbf{\# of} \\\textbf{Activities}} & \tabincell{c}{\textbf{Current \# of} \\\textbf{Deep Links}} & \tabincell{c}{\textbf{\# of}\\ \textbf{Deep Links} \\\textbf{by Aladdin}} & \tabincell{c}{\textbf{Increase of} \\\textbf{Coverage}}\\
  \hline
Cool Wallpapers                & Personalization    & Wallpaper app                     & 5M-10M    & 9   & 0 & 4   & 44\%  \\ \hline
Fitness \& Bodybuilding        & Health \& Fitness  & Fitness \& Bodybuilding           & 1M-5M     & 9   & 0 & 4   & 44\%  \\ \hline
NFL Mobile                     & Sports             & Football app                      & 10M-50M   & 23  & 1 & 12  & 48\%  \\ \hline
Wikipedia                      & Books \& Reference & Wikipedia client                  & 500k-1M   & 18  & 1 & 11  & 56\%  \\ \hline
NPR News                       & News \& Magazines  & News reader                       & 1M-5M     & 17  & 0 & 17  & 100\% \\ \hline
AnkiDroid Flashcards           & Education          & Flash card manager                & 1M-5M     & 21  & 1 & 14  & 62\%  \\ \hline
Lyft Taxi                      & Transportation     & Taxi service                      & 1M-5M     & 12  & 2 & 6   & 33\%  \\ \hline
Booking.com  & Travel \& Local    & Hotel reservations                & 10M-50M   & 120 & 4 & 112 & 90\%  \\ \hline
APUS Booster+                  & Productivity       & System utility                    & 10M-50M   & 20  & 0 & 7   & 35\%  \\ \hline
Hulu                           & Entertainment      & Live streaming                    & 100k-500k & 1   & 0 & 1   & 100\% \\ \hline
Music Player for Android       & Music \& Audio     & Music player                      & 10M-50M   & 13  & 0 & 5   & 38\%  \\ \hline
Marvel Comics                  & Comics             & Marvel Comics                     & 5M-10M    & 33  & 1 & 27  & 79\%  \\ \hline
Free Movies                    & Media \& Vedio     & Movie player                      & 10M-50M   & 1   & 0 & 1   & 100\% \\ \hline
Weather Radar Widget           & Weather            & Weather alert \& forecast         & 1M-5M     & 11  & 0 & 5   & 45\%  \\ \hline
Caller ID    & Communication      & Track phone number location       & 1M-5M     & 11  & 0 & 11  & 100\% \\ \hline
My Diary                       & Lifestyle          & Write Diaries                     & 500K-1M   & 29  & 1 & 26  & 86\%  \\ \hline
Blood Pressure Log   & Medical            & Log blood pressures & 100K-500K & 5   & 0 & 5   & 100\% \\ \hline
Go Fund Me                     & Social             & Personal fundraising platform     & 1M-5M     & 41  & 1 & 13  & 29\%  \\ \hline
Open Camera                    & Photography        & Free camera app                   & 5M-10M    & 2   & 0 & 1   & 50\%  \\ \hline
Timesheet  & Business           & Managing working hours            & 500K-1M   & 26  & 2 & 13  & 42\%  \\ \hline
\end{tabular}
\end{table*}

\subsection{Coverage of Deep Links in Tested Apps}
To investigate the coverage of activities that can be deep linked via Aladdin, we perform static analysis to all the apps in our dataset to derive deep links to activities. We first verify whether the generated deep links can reach the target activities. To this end, we manually collect one instance of each activity to which a deep link is generated, and extract the parameter values in the intents. Then we execute the deep link with needed parameter values and check whether we could reach the same activity instance. Results show that all the deep links can be correctly executed to the target activity instance.

Next, we examine the improvement of deep-link coverage. The total number of activities, the current number of deep links, and the number of deep links by Aladdin are reported from Column 5 to Column 7 in Table~\ref{table:apps}. It can be observed that, although the current number of deep links in these apps is very low (the median is 0), Aladdin can derive deep links for all the apps (with median of 9). In particular, all the activities of 5 apps out of the total 20 apps can have deep links after being analyzed by Aladdin, indicating 100\% coverage.

However, coverage does not reach 100\% in some apps. The main reason is that the intents of some activities encapsulate the objects of app-specific classes, which cannot be resolved by our current solution. We leave this issue for future work.

Considering the improvement of coverage, the median value is 52\%, and and the minimum is 29\%. Therefore, we can conclude that the static analysis of Aladdin can substantially improve the coverage of deep links of existing apps.

\subsection{Accuracy of Dynamic Analysis}
\begin{table}[t]
\centering\scriptsize
\caption{Accuracy of dynamic analysis.}\label{table:accuracy}
\begin{tabular}{llrrrr}
 \hline
  \textbf{App} & \textbf{Activity} & \textbf{Exp.} & \textbf{Result} & \textbf{Recall} & \textbf{Precision}\\
  \hline
  NPR News & SearchActivity & 4 & 4 & 100\% & 100\%\\
  NPR News & NewsStoryActivity & 1 & 1 & 100\% & 100\%\\
  AnkiDroid Flashcards & Statistics & 9 & 5 & 56\% & 100\%\\
  AnkiDroid Flashcards & DeckPicker & 11 & 12 & 100\% & 92\%\\
  APUS Booster+ & HomeActivity & 3 & 3 & 100\% & 100\%\\
  APUS Booster+ & BoostMainActivity & 1 & 4 & 100\% & 25\%\\
  Wikipedia & SettingsActivity & 4 & 4 & 100\% & 100\%\\
  Wikipedia & GalleryActivity & 5 & 5 & 100\% & 100\%\\
  \hline
\end{tabular}
\end{table}
We select 4 apps (NPR News, AnkiDroid Flashcards, APUS Booster+, and Wikipedia) in our dataset to which we can apply dynamic analysis of Aladdin. For each app, we choose 2 activities to find fragments. We also manually examine the number of fragments to get the ground truth for comparison. Table~\ref{table:accuracy} shows the analysis results. Here we use recall and precision to measure the accuracy. The recall is the ratio of fragments correctly found by the dynamic analysis over all the expected fragments. The precision is the ratio of fragments correctly found by the dynamic analysis over all the fragments reported by the dynamic analysis. For 5 out of the 8 activities, the precision and recall are both 100\%, meaning that the dynamic analysis has found all the fragments as expected.

The current dynamic analysis of fragment extraction does have some limitations. In the activity \texttt{Statistics} of \emph{AnkiDroid Flashcards}, the result is smaller than expected. Despite of a 100\% precision, the recall is only 56\%, meaning that developers may have to manually add missing fragments. We find the reason to be that the view which triggers the fragment transition does not have an explicit resource id. To solve this issue in the future, we can use the relative location in the view tree that we build at runtime to identify each view rather than relying on only resource id. On the other hand, in \texttt{DeckPicker} of \emph{AnkiDroid Flashcards} and \texttt{BoostMainActivity} of \emph{AnkiDroid Flashcards}, the recall is 100\% but the precision may not be satisfactory. The reason is the limitation of the structure hash. At runtime, some popping messages and other trivial changes to the original page can result in a total different hash value. As a result, our dynamic analysis should judge the same fragment as a new one. Therefore, taking into account only a single hash value may cause some mistakes in some cases. One possible solution is to record the whole view tree, and design an algorithm to calculate the differences of the view tree after one action. Additionally, only if the difference value reach the threshold can we identify a new fragment.

\subsection{Overhead of Deep-Link Execution}
%\noindent $\bullet$ \textbf{Code size increase}. Aladdin packages the source code of the original app logic together with the code of deep-link proxy and deep-link templates when releasing the \texttt{.apk}. The core logic of the deep-link proxy is \textbf{2,319 lines of Java code} and its size is only 489KB including some third-party libraries. In other words, the size of proxy is rather small compared to that of the original apps. For the size of the deep link templates, we compute the size of the deep link templates for all the 20 apps in our dataset. According to Table~\ref{table:apps}, there are 840 deep link templates generated by Aladdin in total. Figure~\ref{fig:eval:size} shows the distribution of the size. The medium size is 123B, the smallest is 93B, and the largest is 423B. Assume that an app has 1,000 activities, the total code size of templates is no more than 500KB. In other words, Aladdin introduces very tiny code volume compared to the original \texttt{.apk} of the app.

We evaluate the runtime overhead when executing a deep link. Such an overhead comes from executing the replay engine in the deep-link proxy. Note that the proxy works only when a deep link is requested. So there is no overhead when the app is normally used. We use the \texttt{Nexus 5} smartphone model (2.3GHz CPU, 2GB RAM) equipped with Android OS 6 for the evaluation. For each app in the dataset, we select two pages and retrieve their corresponding deep links generated by Aladdin. Then we manually open the two pages for 10 times and execute the deep links for 10 times. At the same time, the average CPU usage and memory consumption of the app are recorded. Evaluation results show that the Aladdin-packaged apps increase no more than 10\% of CPU usage compared to the case that no deep links are executed. Meanwhile, executing deep links requires about 1MB-3MB memory consumption, indicating that the overhead is also quite small. Therefore, the runtime overhead of apps released by Aladdin is very minor and hardly affects the user experiences.
%\begin{figure}[t]
%\centering
%  \includegraphics[width=0.34\textwidth]{figures/template_size}
%  \caption{Distribution of the size of the deep link templates.}~\label{fig:eval:size}
%\end{figure}

\section{Discussion}\label{sec:discussion}
So far, we have described the idea along with the technical details of Aladdin in preceding sections. As the first deep-link releasing tool up to date, we realize that there are some issues worth discussing to validate the real-world applicability of Aladdin.

\noindent $\bullet$ \textbf{Tool limitations}. There are several limitations of Aladdin so that some locations inside apps can not be exposed by deep links via Aladdin. First, Aladdin does not consider the intents with complex object messages so that some activities launched by such kinds of intents cannot be reached. Second, some activities can be launched only in specific states of the app, e.g., after logging in a account. Otherwise, there may be execution errors. So far, Aladdin has not considered such external dependencies of activities. But most apps have implemented complementary logics when the state is not expected. For example, when a user enters an activity that requires a user account but she does not log in, the app may let the user log in first and afterwards return back to the previous activity. Therefore, this limitation is not very significant. Third, there may be pop-up views that has to be closed before performing actions on the activity. Currently, Aladdin does not consider such a condition so that the fragment may be not correctly reached. In the future work, we plan to enhance Aladdin to make it more robust to complex conditions of app executions.

\noindent $\bullet$ \textbf{Developer-desirable release}. The design goal of Aladdin is to help explore all possible locations that can be equipped with deep links. In practice, developers can choose whether the Aladdin-generated deep links are required before releasing their apps. Indeed, just like the web pages, not every location inside an app mandatory needs a deep link. However, Aladdin facilitates developers to more efficiently release their desirable deep links with very little manual efforts. As analyzed previously, releasing deep links can make their apps more user-friendly and embrace more potential collaborators and revenues.

\noindent $\bullet$ \textbf{Instantialization and user-defined.} Aladdin essentially derives deep-link templates and release deep link schemas for Android apps, and thus provides the infrastructural support of wider usage scenario of deep links. An orthogonal issue is how to retrieve the values of parameters to get a concrete deep link. Indeed, such an issue is dependent on the exact context of how deep links are used. When having the deep links, search engines such as Google and Bing can crawl every page of the app, and thus could get all the possible values for the crawled page. It is straightforward to construct the mapping between the actual values and the parameters defined in the schema. App developers can provide additional RESTful APIs to their partners or third-party developers to retrieve the concrete deep links. Since the system can also retrieve all the running information of apps, Android system could leverage the deep-link schema to construct the actual deep links according to the content that a user is now browsing. In this way, it is possible to create ``user-defined'' deep links which the user can bookmark and further share with others, which is a successful experience on the Web. In fact, recent work of In-App Analytic Services~\cite{Suman:HotOS15} has proposed an OS abstraction that can accommodate deep links to support the preceding scenarios.

%\noindent $\bullet$ \textbf{Intent-based transition.} Aladdin essentially relies on the Android's \texttt{Intent} mechanism to distinguish activity transitions. Different instances of intent could transit to different instances of an activity. We also assume that activity transitions can be re-executed by re-sending the intent. However, in practice, some activity transitions do not instantiate different intents. Instead, they use public objects to pass messages. In such a case, the intent instances of activity transitions are always the same so that our approach may not work. One possible solution is to combine taint analysis to find the data related to activity transitions.

\noindent $\bullet$ \textbf{Security and privacy.} Indeed, importing deep links to apps may lead to security and privacy issues. Currently, Aladdin relies on the developers to decide which activities to support deep links or not. For simplicity, we assume that all the instances are permitted to be deep ``linked''. Indeed, deep links provide a mechanism to make the in-app data and content externally ``\textit{accessible}''. As a result,  there might be some potential risks to be attacked by hackers via deep links. However, to keep our contribution clean, addressing the threats of deep links is out of the scope of this paper. In our future work, we plan to extend our preliminary \texttt{WHYPER} technique~\cite{Xie:USENIX2013} with  Aladdin to figure out the possible security and privacy issues.

%Indeed, the idea of deep link is quite new but is showing its importance to break up the barriers of current ``isolated'' apps. Deep link is now developed in a very ad-hoc and tedious way.

\noindent $\bullet$ \textbf{Support of out-of-date content}. Similar to the web hyperlinks, it is argued that the deep links can be out of date and unavailable if the app-content providers remove the content which the deep link refer to~\cite{uLink:MobiSys2016}. As analyzed previously, Aladdin generates deep-link template for the location that the developers desire to be ``linked'', the initialization of deep links are made at runtime. Hence, Aladdin-generated deep links inherently are not affected with respect to the updated or removed content. When the developers decide to add, update, or remove the support of deep links for a location inside their latest released app versions, they need to only run the apps' source code with Aladdin and configure the desirable deep links.

\noindent $\bullet$ \textbf{Generalizability.} Aladdin makes the preliminary effort to facilitate the developers who are willing to release deep links. Although the techniques in this paper is for Android apps that are implemented in Java, the idea and basic principle itself can be extended and applied to other platforms such as iOS and Windows Mobile. Indeed, the static/dynamic analysis techniques over these platforms shall be different.

\noindent $\bullet$ \textbf{Ongoing industrial applicability.} To the best of knowledge, Aladdin makes the first step in releasing deep links with very few developers' manual efforts. %In practice, a top Android app developer ranked by Google Play, called Kika\footnote{Kika. \url{http://www.kika.tech}. Contact the product manager of Kika via \url{ningwang@kikatech.com}.}, is willing to use Aladdin for releasing deep links. Kika has developed a popular keyboard app\footnote{Kika Keyboard. \url{https://play.google.com/store/apps/details?id=com.qisiemoji.inputmethod&hl=en}}. Besides the input method UI, the app has some pages such as in-app ads, widgets, theme store, and so on. As a result, Kika has a strong wish to make users click through to these pages and thus increases potential ads click and revenues. After applying Aladdin, the Kika developers confirmed that most of Aladdin-generated deep links are interesting and valuable. For the next stop, Kika plans to release the deep links to their partners, and checks the actual results over real-world users.
Besides directly applied by app developers, Aladdin can be leveraged as a third-party service in app-centric ecosystem. Based on our previous collaborations~\cite{Li:IMC15, Lu:ICSE16}, Wandoujia is now employing Aladdin to enable the potential collaborations between the apps published on the app store. Aladdin is now under deployment over the Wandoujia's app-analytic platform. When developers submit their apps to Wandoujia, they can optionally choose Aladdin, customize which deep links can be released, and decide which apps can import these deep links. In addition, various famous mobile manufacturers in China, such as \texttt{Huawei} and \texttt{Xiaomi}\footnote{\texttt{Huawei} and \texttt{Xiaomi} are among the top 5 smartphone vendors in China with more than 10 million shipment per quarter. Refer to IDC's report for more details \url{https://www.idc.com/getdoc.jsp?containerId=prCHE41676816}.}, are evaluating Aladdin to derive deep links in their own operating systems, and further support customized intelligent scenarios by ``\textit{mashing-up}'' the app content.

\section{Related Work}\label{sec:related}
In this section, we highlight the literatures related to our work. We first show the industrial and academia efforts on deep links, and then discuss approaches on static and dynamic analysis of Android apps.
\subsection{Deep Link}
Deep link~\cite{deeplink} is an emerging concept for mobile apps. The idea of deep link for mobile apps originates from the links in the Web. It is a uniform resource identifier that links to a specific page inside an app. In the context of Android mobile apps, a URI should be properly handled by configuring intent filters, so as to bring others directly into the specific location within their app with a dedicated link.

Recently, some great companies, especially search engines, have made many efforts on deep links and proposed their criteria for deep links. Google App Indexing~\cite{GoogleAppIndexing} allows people to click from listings in Google's search results into apps on their Android and iOS devices. Bing App Linking ~\cite{BingAppLinking} associates apps with Bing's search results on Windows devices. Facebook App Links~\cite{FacebookAppLinks} is an open cross platform solution for deep linking to content in mobile apps. However, these state-of-the-art solutions all require the need-to-be-deep-linked apps to have corresponding webpages, and the application is narrowed.

The research community is at the early age of studying deep links and very few efforts have been proposed. Azim et al.~\cite{uLink:MobiSys2016} designed and implemented uLink, a lightweight approach to generating user-defined deep links. uLink is implemented as an Android library with which developers can refactor their apps. At runtime, uLink captures intents to pages and actions on each page, and then generates a deep link dynamically, just as bookmarking. Compared to uLink, Aladdin requires smaller developer effort and no intrusive to apps' original code. In addition, Aladdin computes the shortest path to each activity so that we can open a page more quickly than uLink.

Other possible solutions to implement deep links are to leverage the record-and-replay techniques on mobile devices~\cite{RERAN:ICSE2013,Azim:OOPSLA2015}. However, these tools are too heavy-weight~\cite{Flinn:HotMobile2011} and requires either a rooted phone or changes to the mobile OS. So Aladdin provides a developer tool to refactor the apps for deep links, achieving both non-intrusion and light weight execution.

\subsection{Analysis of Inter-Component Communication}
Executing a deep link is highly related to Inter-Component Communication (ICC) of apps. Paulo et al.~\cite{barros2015static} presented static analysis for two types of implicit control flow that frequently appear in Android apps: Java reflection and Android intents. Bastani et al.\cite{bastani2015interactively} proposed a process for producing apps certified to be free of malicious explicit information flows. Li et al.~\cite{IccTA} proposed IccTA to improve the precision of the ICC analysis by propagating context information between components.
%In their approach, the developer provides tests that specify what code is reachable, allowing the static analysis to restrict its search to tested code.
Damien et al.~\cite{Epicc:NDSS2013}\cite{IC3:ICSE2016} developed a tool to analysis the intents as well as entry and exist points among android apps. In their following work~\cite{primo:POPL2016}, they show how to overlay a probabilistic model, trained using domain knowledge, on top of static analysis results, in order to triage static analysis results.
%They apply this idea to analyzing mobile applications. Android application components can communicate with each other, both within single applications and between different applications. In our approach, we apply static code analysis to implement the Navigation Analyzer, which analyzes navigation among activities and builds a Navigation Graph.
%\textbf{Web Crawlers}, also known as Web spiders and (ro)bots, have been studied since the advent of the Web itself~\cite{BrinPage:Google}\cite{Burner1997}\cite{Cho:WWW2001}\cite{Heydon:1999}. There has been extensive research on the hidden Web behind forms~\cite{Barbosa:WWW2007}\cite{Dasgupta:WWW2007}\cite{Fontes:WIDM2004}. The main focus in this research area is to detect ways of accessing the Web content behind data entry points. Alvarez et al.~\cite{Alvarez:2006} discuss some challenges of crawling hidden content generated with JavaScript, but focus on hypertext links. Mesbah et al.~\cite{Mesbah:ICWE2008} was the first academic research work proposing a solution to the problem of crawling AJAX, in the form of algorithms and an open-source tool that automatically crawls and creates a finite state machine of the states and transitions. Our exploitation process brings the idea of Web crawlers, especially crawlers for the hidden Web.
Xu et al.~\cite{WWW17Xu} studied the app collusion which incurs background launching.

\subsection{Automated App Testing}
Aladdin essentially draws lessons from existing app testing efforts~\cite{choi2013guided}\cite{boushehrinejadmoradi2015testing}\cite{DBLP:conf/kbse/ZhangHC15}\cite{Ranorex}\cite{Robotium}, and combines the test inputs generation methodology for the dynamic analysis. Google Android development kit provides two testing tools, Monkey~\cite{Monkey} and MonkeyRunner~\cite{Monkeyrunner}. Hu and Neamtiu~\cite{hu2011gui} developed a useful bug finding and tracing tool based on Monkey. Shauvik et al.~\cite{choudhary2015automated} presented a comparative study of the main existing test input generation techniques and corresponding tools for Android. Ravi et al.~\cite{bhoraskar2014brahmastra} presented an app automation tool called Brahmastra to the problem of third-party component integration testing at scale
% in which one party wishes to test a large number of applications using the same third-party component for a potential vulnerability.
Machiry et al.~\cite{machiry2013dynodroid} presented a practical system Dynodroid for generating relevant inputs to mobile apps on the dominant Android platform. It uses a novel ``observe-select-execute'' principle to efficiently generate a sequence of such inputs to an app. %These research work mainly focuses on automated test inputs generation.
Azim et al.~\cite{azim2013targeted} presented A3E, an approach and tool that allows substantial Android apps to be explored systematically while running on actual phones. The key insight of their approach is to use a static, taint-style, dataflow analysis on the app bytecode in a novel way, to construct a high-level control flow graph that captures legal transitions among activities (app screens). Hao et al.~\cite{Hao:MobiSys2014} designed PUMA, a programmable UI automation framework for conducting
dynamic analyses of mobile apps at scale. PUMA incorporates a generic monkey and exposes an event driven programming abstraction. Different from these tools and systems, the goal of dynamic analysis in Aladdin is to identify fragments or sub-screens of activities that are internal states of apps. So we design UI-tree-based fragment identification and fragment transition graph to address such an issue.

% and app systematically exploration for automated deep links generation.

%In industry, Monkey is an automated fuzz testing tool creates random inputs without considering application's state.

%MonkeyRunner is a remote testing framework. A user can control application running on the phone from the computer through USB connection. There are also some other famous automated software testing tools, such as Ranorex~. We adopt Robotium to stimulate click operation on Android phones.

\section{Conclusion}\label{sec:conclusion}
In this paper, we have presented an empirical study of deep links on 20,000 Android apps. Although more and more apps have supported deep links, the coverage of deep links is still very low and implementing deep links requires non-trivial developer manual efforts. To address the issues, we propose Aladdin, a novel approach to automatically release deep links for Android apps. Leveraging the static analysis and dynamic analysis of Android apps, Aladdin can derive deep links to arbitrary locations inside an app. Aladdin integrates a deep-link proxy together with the original source code of the app to realize executing deep links at runtime. So there are no changes to the original business logic and no coding effort from developers. The evaluations on 20 apps demonstrate that the coverage of deep links can be increased by 52\% on average while incurring minimal runtime overhead.

Some ongoing efforts shall make Aladdin more practical. First, we are enhancing the static analysis of activities to resolve the objects of app-specific classes encapsulated in the intents to further improve the coverage. Second, we are optimizing the algorithm of dynamic analysis to improve the precision and recall of fragments, and to make the analysis more efficient. Last but not the least, we plan to apply Aladdin to open-source apps as well as commercial apps in order to get feedbacks from app developers and evaluate Aladdin in the real world.

\bibliographystyle{abbrv}
\bibliography{inapp,mobisaas}
\end{document}